\documentclass[12pt]{article}
\usepackage[margin=1in]{geometry}

\usepackage{subfigure}
\usepackage{epsfig, graphics, cite}
\usepackage{amsmath,amssymb}
\usepackage{times}
\usepackage{url}
\usepackage[breaklinks=true,filecolor=black,citecolor=black,urlcolor=black,linkcolor=black,colorlinks,pdfpagelabels,pdfpagelayout=SinglePage,dvips]{hyperref}
\usepackage[hyphenbreaks]{breakurl}
\usepackage{balance}
\usepackage{multirow}
\usepackage{breakurl}

\newcommand{\eg}{{\it e.g.}}
\newcommand{\ie}{{\it i.e.}}

\usepackage{color}
\definecolor{orange}{RGB}{153,153,255}

\newcommand{\para}[1]{\vspace{1mm}\noindent{\bf #1}}

\newenvironment{pitem}{
 \begin{itemize}
 \setlength{\itemsep}{1pt}
 \setlength{\parskip}{0pt}
 \setlength{\parsep}{0pt}
}{\end{itemize}}

\title{Modeling Tiered Pricing in the Internet\\ Transit Market}

\author{
Vytautas~Valancius, Cristian~Lumezanu, Nick~Feamster,\\
Ramesh~Johari, and~Vijay~V.~Vazirani%
\thanks{V.~Valancius, N.~Feamster, and V.V.~Vazirani are with the Georgia
Institute of Technology. C.~Lumezanu is with NEC Labs USA. R.~Johari is with Stanford University.}}

\begin{document}
\maketitle

\begin{sloppypar}
\begin{abstract}

ISPs are increasingly selling ``tiered'' contracts, which offer Internet
connectivity to wholesale customers in bundles, at rates based on the
cost of the links that the traffic in the bundle is traversing. Although
providers have already begun to implement and deploy tiered pricing
contracts, little is known about how such pricing affects ISPs and their
customers. While contracts that sell connectivity on finer granularities
improve market efficiency, they are also more costly for ISPs to
implement and more difficult for customers to understand.  In this work
we present two contributions: (1)~we develop a novel way of mapping
traffic and topology data to a demand and cost model; and (2)~we fit this
model on three large real-world networks: an European transit ISP, a
content distribution network, and an academic research network, and run
counterfactuals to evaluate the effects of different pricing strategies
on both the ISP profit and the consumer surplus. We highlight three core
findings. First, ISPs gain most of the profits with {\em only three or
four pricing tiers} and likely have little incentive to increase
granularity of pricing even further. Second, we show that {\em consumer
surplus follows closely, if not precisely, the increases in ISP profit}
with more pricing tiers. Finally, the common ISP practice of structuring
tiered contracts according to the cost of carrying the traffic flows
(\eg{}, offering a discount for traffic that is local) can be suboptimal
and that dividing contracts based on {\em both traffic demand and the
cost of carrying it} into {\em only three or four tiers} yields
near-optimal profit for the ISP.

\end{abstract}

\section{Introduction}

The increasing commoditization of Internet transit is changing the
landscape of the Internet bandwidth market. Although residential
Internet Service Providers (ISPs) and content providers are connecting
directly to one another more often, they must still use major Internet
transit providers to reach most destinations. These Internet transit
customers can often select from among dozens of possible
providers~\cite{peeringdb}. As major ISPs compete with one another, the
price of Internet transit continues to plummet: on average, transit
prices are falling by about 30\% per year~\cite{Norton-www}.

As a result of such competition, ISPs are evolving their business models
and selling transit to their customers in many ways to try to retain
profits. In particular, many transit ISPs today are increasingly
implementing pricing strategies where wholesale Internet transit is
priced by volume or destination. For example, ISPs charge prices on
traffic bundles based on factors, such as how far the traffic is
traveling, and whether the traffic is ``on net'' (\ie, to that ISP's
customers) or ``off net''~\cite{www-guavus-tiers}.  Still, we understand
very little about the efficiency of such pricing schemes. In this work,
we study {\em destination-based tiered pricing}, with the goal of
understanding how tiered pricing affects ISP profit and consumer
surplus.

Although understanding the effects of different transit pricing
structures is important, modeling them is quite difficult. The model must
take as an input existing customer demand and predict how traffic (and,
hence, ISP profit and consumer surplus) would change in response to
pricing strategies. Such a model must capture how customers would
respond to any pricing change---for any particular traffic flow---as well
as the change in cost of forwarding traffic on various paths in an ISP's
network. Of course, many of these input values are difficult to come by
even for network operators, but they are especially elusive for
researchers; additionally, even if certain values such as costs are
known, they change quickly and differ widely across ISPs.

The model we develop allows us to estimate the relative effects of tiered
pricing scenarios, despite the lack of availability of precise values for
many of these parameters. The general approach, which we describe in
Section~\ref{sec:model}, is to start with a demand and cost model and
assume both that ISPs are already profit-maximizing and that the current
prices reflects both customer demand and the underlying network costs.
These assumptions allow us to either fix or solve for many of the unknown
parameters and run counterfactuals to evaluate the relative effects of
dividing the customer demand into pricing tiers. To drive this model, we
use traffic data from three real-world networks: a major international
content distribution network with its own network infrastructure; an
European transit ISP; and an academic research network. We map the demand
and topology data from these networks to a model that reflects the
service offerings that real-world ISPs use.

Using our model and the datasets, we evaluate how the tiered
pricing impacts ISPs and consumers. According to economic theory
prescriptions, as we increase pricing granularity, we would expect to
observe two trends: 1) ISPs should increase their profit with more tiers
and 2) the marginal profit increase should be diminishing with the
number of tiers. As we confirm (or refute) these predictions, we also
hope to find answer to these two questions:

\begin{pitem}
  
\item {\em How many pricing tiers ISPs need to introduce to maximize
  their profit?} ISPs today usually use 2 to 6 pricing tiers. If, for
  instance, we show that a greater number of tiers provides a tangible
  benefit, the ISPs could rethink their current pricing policies.

\item {\em How does consumer surplus change with an increasing number
  pricing tiers?} In abstract, a more granular pricing should lead to a
  better resource allocation efficiency, thus increasing the consumer
  surplus.  If, on the other hand, we find a significant drop in consumer
  surplus, policy makers could take the results of such modeling into
  account when reviewing regulations of the Internet transit market.
  
\item {\em What are the best ways for ISPs to structure the pricing
  tiers?} When ISPs use limited number of pricing tiers, they have do
  decide which destinations to bundle together for uniform pricing.
  Today, ISPs, when they use tiered pricing, usually group destinations
  based on cost (\eg{}, pricing local traffic cheaper than global
  traffic). We will evaluate different destination bundling strategies to
  see if the current approach is adequate.

\end{pitem}

\noindent As we look for answer to these questions, we make the following
four contributions. First, to analyze the effects of tiered pricing, we
develop a model that captures demands and costs in the transit market.
One of the challenges in developing such a model is applying it to real
traffic data, given many unknown parameters (\eg, the cost of various
resources, or how users respond to price). Hence, we devise methods for
fitting empirical traffic demands to theoretical cost and demand models
(Section~\ref{sec:model}). Second, we apply our model to real-world
traffic matrices and network topologies to characterize the effects of
tiered pricing on the ISP profit and consumer surplus. Third, we evaluate
six algorithms to structure the pricing tiers (Section~\ref{sec:eval}).
Fourth, we perform robustness analysis of our model, to see if results
hold for a range of input parameters (Section~\ref{sec:sensitivity}). We
wrap up the work with a review of related work and conclusion
(Sections~\ref{sec:related} and~\ref{sec:conclusion}). Before we start,
however, in the next section we present the overview the examples of
tiered pricing and explain why is it beneficial to ISPs.

\section{Background}
\label{sec:background}

In this section, we describe the current state of affairs in the Internet transit
market. We first taxonomize {\em what} services (bundles) ISPs are
selling. We then provide intuition on {\em why} ISPs are moving towards
tiered wholesale Internet transit service.

\subsection{Current Transit Market Offerings}
\label{sec:services}

Unfortunately, there is not much public information about the wholesale
Internet transit market. ISPs are reluctant to reveal specifics about
their business models and pricing strategies to their competitors.
Therefore, to obtain most of the information in this section, we engaged in
many discussions and email exchanges with network operators.  Below, we
classify the types of Internet transit service we identified during
these conversations.  Although much of the information in this section is
widely known in the network operations community, it is difficult to find
a concise taxonomy of product offerings in the wholesale transit market.
The taxonomy below serves as a point of reference for our discussions of
tiered pricing in this paper, but it may also be useful for anyone who
wishes to better understand the state of the art in pricing strategies in
the wholesale transit market.

\para{Peering} business relationships (or, more formally, {\em
settlement-free peering} relationships) have been extensively studied by
networking researchers~\cite{Johari2003, Chang:infocom2006,
Dhamdhere2010, Feamster2004b, Feamster2004c} and well-documented in the
industry white papers~\cite{Norton-www}. Most peering connections are
established through public Internet eXchange Points (IXP), while higher
bandwidth peering often requires private peering sessions. A network
engaging in a settlement free peering allows its peer to reach {\em
on-net} destinations---destinations in its own network, and destinations
in customer networks. For the peering to be settlement-free, most ISPs
pose a set of requirements to prospective peers, such as number of
interconnection points, geographic coverage, or ingress/egress traffic
ratios. If an ISP cannot meet peering requirements, it is forced to buy
Internet {\em transit} or {\em paid peering}~\cite{l3-cogent-depeering,
telia-cogent-depeering, www-he-cogent-depeering}.

\para{Transit.} Most ISPs offer conventional Internet transit service.
Internet transit is sold at a {\em blended rate}---a single price
(usually expressed in \$/Mbps/month)---charged for traffic to all
destinations. Historically, blended rates have been decreasing by 30\%
each year~\cite{Norton-www}. Blended rate is the simplest and yet the
most crude way to charge for traffic. If network costs are highly
variable, less costly flows in the blended-rate bundle subsidize other,
more expensive flows. ISPs often innovate by offering more than one rate:
We summarize three pricing models that require two or more rates:
(1)~{\em paid peering}, (2)~{\em backplane peering}, and (3)~{\em
regional pricing}.

\para{Paid peering} is similar to settlement-free peering, except that
one network pays to reach the other. A major ISP might separately sell
{\em off-net} routes (wholesale transit) at one rate and {\em on-net}
routes (to reach destinations inside its own network) at another (usually
lower) rate. For example, national ISPs in Eastern Europe, Australia, and
in other regions may sell local connectivity at a discount to increase
demand for local traffic, which is is significantly cheaper than transit
to outside global destinations~\cite{www-adam-unmetered}.  The on-net
routes are also offered at a discount by some major transit ISPs to large
content providers, because such transit ISPs can recoup part of the costs
from their customers, who congest paid upstream links to transit ISPs by
downloading the content. Some instances of paid peering have spawned
significant controversy: most recently, Comcast---primarily a network
serving end-users---was accused of a network neutrality violation when it
forced one tier-1 provider to pay to reach Comcast's
customers~\cite{www-comcast-vs-level3}.

\para{Backplane peering} occurs when an ISP, in addition to selling
global transit through its own backbone, charges a discount rate for the
traffic it can offload to its peers at the same Internet exchange.
Smaller ISPs buy such a service because they might not meet all the
settlement-free peering requirements to peer directly with the ISPs in
the exchange. Although many large ISPs discourage this practice, some
ISPs deviate by offering backplane peering to retain customers or to
maintain traffic ratios with their peers. As with paid peering, the ISP
selling backplane peering has to account and charge for at least two
traffic flows: one to peers and another to its backbone.

\para{Regional pricing} occurs when transit service providers offer
different rates to reach different geographic regions. The regions can be
defined at different levels of granularity, such as PoP, metro
area, regional area, nation, or continent. In some instances, the
transit ISP offers access to all regions with different prices; in other
instances, the downstream network purchases access only to a specific
geographic region (\eg{}, access only to South America or Australia). In
practice, due to the overhead of provisioning and maintaining many
sessions to the same customer, ISPs rarely use more than one or two extra
price levels for different regions.

We speculate that the bundling strategies described above arose primarily
from operational and cost considerations. For example, it is relatively
easy for a transit ISP to tag which routes are coming from customers and
which routes are coming from peers and then in turn sell them separately
to its customers. Similarly, it is relatively easy to sell local (\ie,
less costly) routes separately. We show that these na\"{i}ve bundling
strategies might not be as effective as bundling strategies that account
for both cost and demand.

\subsection{The Trend Towards Tiered Pricing}
\label{sec:motivation}

Conventional blended rate pricing is simple to implement, but it may be
inefficient.  ISPs can lose profit as a result of blended-rate pricing,
and customers can lose surplus.  This is an example of {\em market
failure}, where goods are not being efficiently allocated between
participants of the market.  Another outcome of blended-rate pricing is
the {\em increase in direct peering} to circumvent ``one-size-fits-all''
transit.  Both phenomena provide incentives for ISPs to improve their
business models to retain revenue.  We now explain each of these
outcomes.

\begin{figure}[t!]
  \centering
  \subfigure[{\bf Blended-rate pricing.} ISP charges a single blended
  rate $P_0$.]{
  \includegraphics[width=0.7\linewidth]{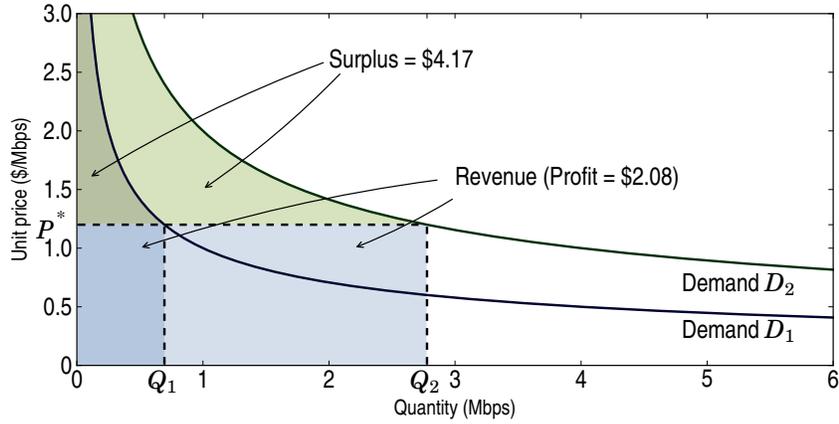}
  \label{fig:bundled_pricing}}
  \subfigure[{\bf Tiered pricing.} ISP charges rates $P_1$ and $P_2$ for flows.]{
  \includegraphics[width=0.7\linewidth]{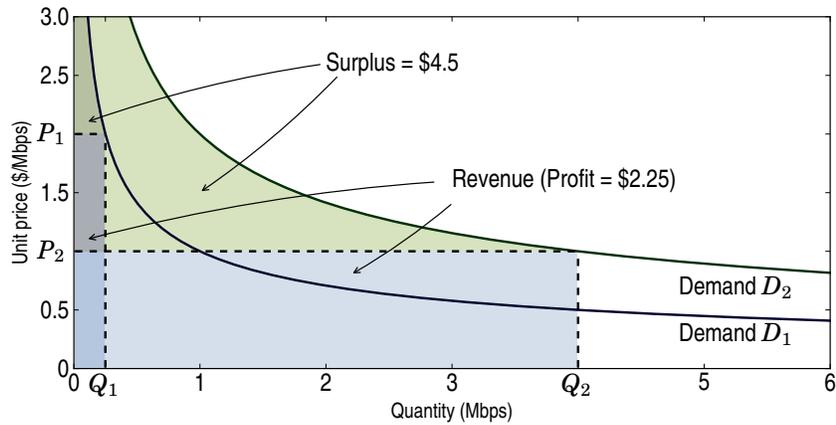}
  \label{fig:unbundled_pricing}}
  \centering
  \caption{Market efficiency loss due to coarse bundling.}
  \label{fig:inefficiency}
\end{figure}

\subsubsection{Profit and surplus loss} 

Selling transit at a blended rate could reduce profit for transit ISPs
and surplus for customers. We define an ISP's {\em profit} as its revenue
minus its costs, and {\em customer surplus} as customer utility minus the
amount it pays to the ISP. Unrealized profit and surplus can occur when
ISPs charge a single rate while incurring different costs when delivering
traffic to destinations.


Figure~\ref{fig:inefficiency} illustrates how tiered pricing can increase
both the profit for an ISP and the surplus for a customer. The
downward-sloping curves represent consumer demand\footnote{We model
consumer demand as {\em residual demand}. Residual demand  accounts for
consumption change both due to inherent consumer demand and due to some
consumers shifting consumption to substitutes, such as other ISPs (See
Section~\ref{sec:ced}.)} to two destinations.  Since the demand slope
$D_2$ is higher than demand slope $D_1$, the customer has higher
demand for the second destination in the ISP's network. Assume that the
ISP cost of serving demand $D_1$ is \$1, while the cost of serving demand
$D_2$ is \$0.5.  Modeling demand with constant elasticity
(Section~\ref{sec:model}), the {\em profit maximizing price} can be shown
to be $P_0=\$1.2/Mbps$.  If, however, the ISP is able to offer two
bundles, then the profit maximizing prices for such bundles would be
$P_1=\$2.7$ and $P_2=\$1$.  Figure~\ref{fig:unbundled_pricing} shows that
this price setup not only increases ISP profit but also increases
consumer surplus and thus social welfare. 

The market achieves higher efficiency because customers adjust their
consumption levels of the ISP network according to their demand and to
the prices that the ISP exposes, which directly depend on its costs.
Without the ISP's indirect exposure of its costs, the customer consumes
less of the cheaper capacity and more of the expensive capacity than it
would otherwise. In Section~\ref{sec:model}, we formalize the market that
we have used in this example and propose more complex demand and cost
models.

\begin{figure}[t!]
  \centering
  \includegraphics[width=0.6\linewidth]{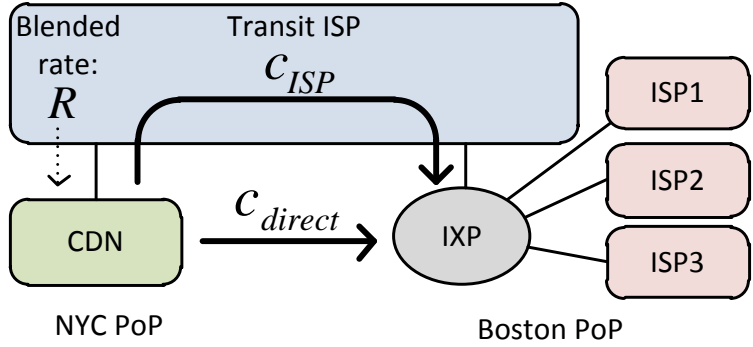}

  \caption{The customer procures a direct link if the
  cost for such a link is lower than the blended rate  $c_{direct} <  R$.}
  
  \label{fig:cdn-game} 
\end{figure}

\subsubsection{Increase in direct peering} 

Charging for traffic at a blended rate also provides incentives for
client networks to connect directly to georgraphically close Internet
Exchange Points (IXPs). For
instance, if a transit ISP charges only blended rate, client network
might find geographically close IXPs cheaper to reach by leasing or
purchasing private links. While direct peering is generally perceived as
a positive phenomenon, for transit ISPs it means less revenue. Direct
peering efforts can also diminish economies of scale: instead of using
shared ISP infrastructure, customers provision their own connectivity to
IXPs. 

Figure~\ref{fig:cdn-game} illustrates an interaction between an upstream
ISP and a CDN client (\eg{}, Google, Microsoft) with its own
backbone, which extends to the NYC PoP. The CDN might, or might not, have
a content cache at the Boston IXP, but since it does not have its own
backbone presence at the IXP, the CDN must pay the upstream ISP to reach
it.  The ISP offers a blended rate $R$ at the the NYC PoP for all the
traffic, including the traffic to the Boston IXP. The blended rate $R$ is
set to compensate the upstream provider for the overall traffic mix and,
therefore, is {\em higher} than the amortized cost of most of the cheaper
(more localized) flows that ISP is serving (\ie{}, the flows between the
NYC and Boston PoPs). The CDN eventually will procure a direct link to
the Boston IXP, if it finds that it can procure such a direct link at an
amortized cost $c_{direct} < R$.  Assuming the ISP's profit margin is $M$
and flow accounting overhead is $A$ (discussed in
Section~\ref{sec:accounting}), such a direct link presents a {\em market
failure} if $c_{direct} > (M+1)c_{ISP} + A$, because the customer deploys
additional capacity at a higher cost than the ISP could have charged in a
tiered market.

Some operators we interviewed confirm that they periodically re-evaluate
transit bills and expand their backbone coverage if they find that having
own presence in an IXP pays off. In today's transit market, many
customers increasingly opt for direct
peering~\cite{Labovitz:sigcomm2010}; transit service providers are
absorbing losses as a result of competitive pressure~\cite{Norton-www}.
Naturally, this pressure increases the incentive for ISPs to adopt a
tiered pricing model for local traffic. The central question, then, is
how they should go about structuring these tiers. The rest of the paper
focuses on this question.

\section{Modeling Profits, Costs, and\\ Demands}
\label{sec:model}

We develop demand and cost models that capture ISP profit under various
pricing strategies. Since no model can perfectly capture demand in the
Internet transit market, we perform our evaluation with two commonly used
demand models. Because cost is also difficult to model, we devise four
network cost models. We first define ISP profit and then describe demand
and cost models.

\vspace{-3mm}
\subsection{ISP Profit}

We consider a transit market with multiple ISPs and customers.  Each ISP
is rational and maximizes its profit, which we express as the difference
between its revenue and costs:

\begin{equation}
  \Pi(\vec{P}) =
  \textstyle\sum_{i=1}^{n}\left(p_iQ_i(\vec{P}) -
  c_iQ_i(\vec{P})\right)
  \label{eqn:profit}
\end{equation}

\noindent where $n$ is the number of flows ISP is serving,
$p_i\in\vec{P}$ is the price an ISP sets to deliver flow $i$, $c_i$ is
the unit cost for $i$, and $Q_i(\vec{P})=q_i$ is the demand for $i$ given a
vector of prices $\vec{P} = (p_1, p_2, \ldots, p_n)$. An ISP chooses the
price vector $\vec{P}$ that maximizes its profit (in case when ISP is
using only single pricing tier, the prices are effectively set to
$p_1=p_2\cdots=p_n$).

Given knowledge of both the traffic demand of customers and the costs
associated with delivering each flow, we can compute ISP profit.
Unfortunately, it is difficult to validate any particular demand function
or cost model; even if validation were possible, it is likely that cost
structures and customer demand could change or evolve over time.
Accordingly, we evaluate ISP profit for various tiered pricing approaches
under a variety of demand functions and cost models.
Section~\ref{sec:demand-model} describes the demand functions that we
explore, and Section~\ref{sec:cost-model} describes the cost models that
we consider.

\vspace{-3mm}
\subsection{Customer Demand}
\label{sec:demand-model}

To compute ISP profit for each pricing scenario, we must understand how
customers adjust their traffic demand in response to price changes. We
consider two families of demand functions: {\em constant elasticity} and
{\em logit}.  

\para{Constant elasticity demand.} The constant elasticity demand (CED)
is derived from the well-known {\em alpha-fair} utility
model~\cite{Mo:2000}, which is often used to model user utility on the
Internet. The alpha-fair utility takes the form of a concave increasing
utility function, which emulates a decreasing marginal benefit to
additional bandwidth for a user. In this model flow demands are {\em
separable} (\ie{}, changes in demand or prices for one flow have no
effect on demand and prices of other flows). The CED model is most
appropriate for scenarios when consumers have no alternatives (\eg{},
when the content that a customer is trying to reach is not replicated, or
the customer needs to communicate with a specific endpoint on the
network).

\para{Logit demand.} To capture the fact that customers might sometimes
have a choice between flows (\eg{}, sending traffic to alternative
destination if the current one becomes too expensive), we also perform
our analysis using the {\em logit model}, where demands are not
separable: the price and demand for any flow depend on prices and demands
for the other flows.  The logit model is frequently used for this purpose
in econometric demand estimation \cite{McFadden1973}.  In the logit
model, each consumer nominally prefers the flows that offers the highest
utility. This matches well with scenarios when consumers have several
alternatives (\eg{}, when requested content is replicated in multiple
places).

\subsubsection{Constant elasticity demand}
\label{sec:ced}

\begin{figure}[t]
\centering
\includegraphics[width=0.7\linewidth]{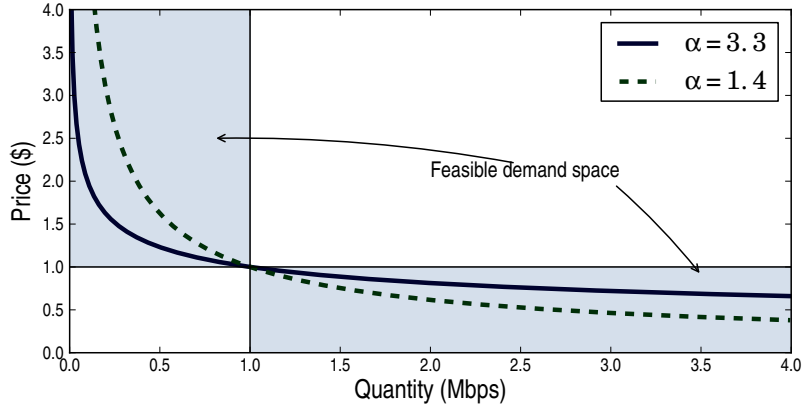}
\caption{Feasible CED demand functions for $v=1$.}
\label{fig:demand}
\end{figure}

The CED demand function is defined as follows:
\begin{equation}
  Q_i(p_i) = \left(\frac{v_i}{p_i}\right)^{\alpha}
  \label{eqn:demand-ced}
\end{equation}

\noindent where $p_i$ is the unit price (\eg{}, \$/Mbit/s), $\alpha \in
(1,\infty)$ is the price sensitivity, and $v_i>0$ is the valuation
coefficient of flow $i$. The demand function can be interpreted to
represent either {\em inherent} consumer demand or {\em residual}
consumer demand, which reflects not only the inherent demand but also the
availability of substitutes.

Figure~\ref{fig:demand} presents example CED demand functions for $v=1$
and two values of $\alpha$, $3.3$ and $1.4$. Higher values of $\alpha$
indicate high elasticity (users reduce use even due to small changes in
price).  For example, the demand with elasticity $\alpha=3.3$ might
represent the traffic from residential ISPs, who are more sensitive to
wholesale Internet prices and who respond to price changes in a more
dramatic way.  Similarly, the demand with elasticity $\alpha=1.4$ might
represent the traffic from enterprise customers, who are less sensitive
to the Internet transit price changes.  Although our model does not
capture full dynamic interaction between competing ISPs (\eg{}, price
wars), modeling demand as residual allows us to account for the existing
competitive environment and switching costs.  As discussed above, high
elasticity can also indicate that competitors are offering more
affordable substitutes, and that switching costs for customers are low.
In our evaluation, we use a range of price sensitivity values to measure
how ISP profit changes for different values of the elasticity of user
demand. The gray area in Figure~\ref{fig:demand} shows that we can cover
all feasible demand functions simply by varying $\alpha$.

\para{CED profit.} Using the expressions for ISP profit
(Equation~\ref{eqn:profit}) and demand (Equation~\ref{eqn:demand-ced}),
and assuming separability of demand of different flows, the ISP profit is:

\begin{equation}
  \label{eqn:profit-ced}
  \Pi(\vec{P}) = \textstyle\sum_{i=1}^{n}\left(\frac{v_i}{p_i}\right)^{\alpha}\left(p_i
  - c_i\right).
\end{equation}

\para{CED profit-maximizing price.} By differentiating the profit, we find
the profit-maximizing price for each flow $i$:

\begin{equation}
 p_i^* = \textstyle\frac{\alpha c_i}{\alpha-1}.
 \label{eqn:ced-optimal-price}
\end{equation}

\para{CED consumer surplus.} Consumer surplus is the difference between
consumer utility and the price paid. Price at the equilibrium is equal to the
marginal utility, and thus we can find utility by integrating price as a
function of demand ($p_i=v_i/q_i^{1/\alpha}$,
Equation~\ref{eqn:demand-ced}) in terms of $q_i$. Substituting in the
resulting equation quantity with price and substracting the price
paid, we get consumer surplus expression as function of price:

\begin{equation}
  CS(\vec{P}) =
  \textstyle\sum_{i=1}^n\left(\frac{\alpha
  v_i^{\alpha}p_i^{1-\alpha}}{\alpha-1} - p_i \right)
\end{equation}

\begin{figure}
  \centering
  \includegraphics[width=0.7\linewidth]{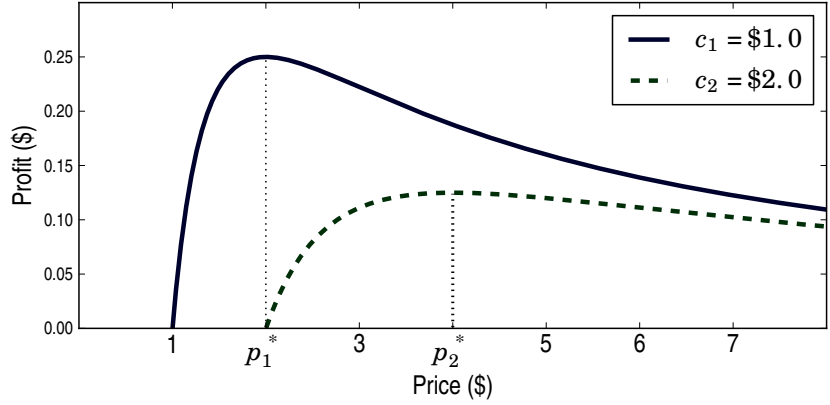}
  \caption{Profit for two flows with identical
  demand ($v_1=v_2=1.0$, $\alpha=2$) but different cost.}
  \label{fig:profit-ced}
\end{figure}

Figure~\ref{fig:profit-ced} illustrates profit maximization for two flows
that have identical demand functions but different costs.  For example,
the first flow costs $c_1=\$1.0$ per unit to deliver and mandates optimal
price $p^*=\$2.0$ which results in $\$0.25$ profit. The second flow is
more costly thus the profit maximizing price is higher. In this case, the
first plot might represent profit for local traffic, while the second
plot represents national traffic: ISPs must price national traffic
higher than local-area traffic to maximize profit. 

\para{CED price for bundled flows.} In our evaluation, we test various pricing strategies that bundle
multiple flows under the same profit-maximizing price. To find the price for each
bundle, we first map real world demands to our model to obtain the
valuation $v_i$ and cost $c_i$ for each flow. Then, we differentiate the
profit (Equation~\ref{eqn:profit-ced}) with respect to the price of each
bundle. For example, when we have a single bundle for all flows,
we obtain the following profit-maximizing price:

\begin{equation}
  P^* = \frac{\alpha\sum_{i=1}^n c_iv_i^{\alpha}}{(\alpha-1)\sum_{i=1}^n
  v_i^{\alpha}}
  \label{eqn:ced-optimal-price}
\end{equation}

\noindent where $n$ is the number of flows. Section~\ref{sec:eval}
details this approach.

\subsubsection{Logit demand}
\label{sec:logit}

The logit demand model assumes that each consumer faces a discrete choice
among a set of available goods or services. In the context of data
transit, the choice is between different destinations or flows. Following
Besanko {\em et al.}~\cite{Besanko:logit1998}, a consumer $j$ using flow
$i$ will obtain the surplus:

\begin{equation*}
u_{ij}=\alpha(v_i - p_i) + \epsilon_{ij}
\end{equation*}

\noindent where $\alpha\in(0,\infty)$ is the elasticity parameter, $v_i$
is the ``average'' consumer's {\em maximum willingness to pay} for flow
$i$, $p_i$ is a price of using $i$, and $\epsilon_{ij}$ represents
consumer $j$'s idiosyncratic preference for $i$ (where $\epsilon_{ij}$
follows a Gumbel distribution.) The elasticity parameter in logit demand
model serves similar function to elasticity parameter in constant
elasticity demand model: high $\alpha$ values represnt high elasticity
(or abundance of competition), while low $\alpha$ values represent lower
elasticity. The logit model defines the probability that any given
consumer will use flow $i$ as a function of the price vector of all
flows:

\begin{equation}
  \textstyle s_i(\vec{P}) = \frac{e^{\alpha(v_i-p_i)}}{\sum_{j=1}^n e^{\alpha(v_j-p_j)} + 1}
 \label{eqn:marketshare}
\end{equation}

\noindent where $\sum_{i=0}^n s_i(\vec{P}) = 1$. The demand for flow $i$
equals the product of $s_i(\vec{P})$ and the total number of
consumers ($K$):

\begin{equation}
 \textstyle Q_i(\vec{P}) = K s_i(\vec{P}).
 \label{eqn:demand-logit}
\end{equation}

\noindent Here, $s_i$ is also called the {\em market share} of flow $i$. The
model also accounts for the possibility that some customers elect not to
send traffic to any destination. The market share for traffic not sent
is:

\begin{equation}
  \textstyle s_0(\vec{P}) = \frac{1}{\sum_{j=1}^n e^{\alpha(v_j-p_j)} + 1}.
  \label{eqn:share-zero}
\end{equation}

\begin{figure}
\centering
\includegraphics[width=0.7\linewidth]{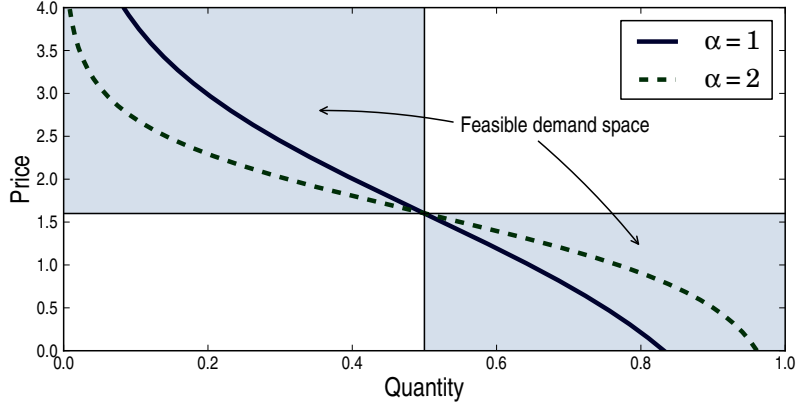}
\caption{Logit demand function.}
\label{fig:logit-demand}
\end{figure}

Figure~\ref{fig:logit-demand} shows examples of logit demand functions.
We assume a setting with two flows, with two values for the
valuation $v_i$, $1.6$ and $1$.  We fix the price for the first
flow to $1$, and we vary the price for the second flow
between $0$ and $4$. The figure shows demand curves for the second
flow, for two values of $\alpha$.  Similar to the constant
elasticity demand model, lower values of $\alpha$ indicate low elasticity
of demand, where users need bigger price variations to modify their
usage.

\para{Logit profit.} Using the expressions for ISP profit
(Equation~\ref{eqn:profit}) and logit demand
(Equation~\ref{eqn:demand-logit}), the ISP profit is:

\begin{equation}
  \Pi(\vec{P}) = \textstyle K\sum_{i=1}^n s_i(\vec{P}) (p_i - c_i).
\label{eqn:logit-profit}
\end{equation}

\para{Logit profit-maximizing prices.} To find the profit maximizing price
for flow $i$, we find the first-order conditions for
Equation~\ref{eqn:logit-profit}:

\begin{equation}
p_i^* = c_i + \frac{1}{\alpha s_0}.
\label{eqn:logitprofit}
\end{equation}

\noindent Due to the presence of $s_0$, $p_i^*$ recursively depends on
itself and on profit-maximizing prices of other flows. To obtain maximum
profit, we develop an iterative heuristic based on gradient descent that starts from
a fixed set of prices ($p_i=P_0, \forall i$) and greedily updates them towards the optimum.

\para{Logit consumer surplus.} After ISP sets profit-maximizing prices
$\vec{P}$, we can compute consumer surplus. We find consumer surplus
expression by taking the expectation of the sum of all the consumer
utilities:

\begin{equation}
  \textstyle CS(\vec{P}) = K\textstyle\frac{\gamma + \ln(\textstyle\sum_{i=1}^n
  e^{\alpha(v_i - p_i)}+1)}{\alpha}
\end{equation}

\noindent where $\gamma$ is Euler's constant~\cite{Ali08}.

\para{Valuation and cost of bundled flows.} To test pricing strategies,
we first map real traffic demands to the model to find the valuation
$v_i$ and cost $c_i$ for each flow $i$. We then bundle the flows as
described in Section~\ref{sec:bundling-strategies}. Knowing that $\sum_i
s_i = 1$ and applying Equation~\ref{eqn:marketshare} allows us to
compute valuations 
for any bundle of flows as:

\begin{equation}
  v_{bundle} = \frac{\ln\left(\sum_{i=1}^n e^{\alpha v_i}\right)}{\alpha}
\end{equation}

\noindent where $v_i$ are valuations of the flows in the bundle.
Similarly we can find the average unit cost of combined flows in each bundle:

\begin{equation}
  c_{bundle} = \frac{\sum_{i=1}^n c_ie^{\alpha v_i}}{\sum_{i=1}^n
  e^{\alpha v_i}}.
\end{equation}

\vspace{-3mm}
\subsection{ISP Cost}
\label{sec:cost-model}

Modeling cost is difficult: ISPs typically do not publish the details of
operational costs; even if they did, many of these figures change rapidly
and are specific to the ISP, the region, and other factors. To account
for these uncertainties, we evaluate our results in the context of
several cost models. We also make the following assumptions.  First, we
assume the more traffic the ISP carries, the higher cost it incurs. Although
on a small scale the bandwidth cost is a step function (the capacity is
added at discrete increments), on a larger scale we model cost as a
linear function of bandwidth. Second, we assume that ISP transit cost
changes with distance.  
Both assumptions are motivated by practice:
looking only at specific instances of connectivity, the cost is a step
function of distance (\eg, equipment manufacturers sell several classes
of optical transceivers, where each more powerful transceiver able to
reach longer distances costs progressively more than less powerful
transceivers~\cite{www-cisco-transceivers}). Over a large set of links,
we can model cost as a smooth function of distance.

The cost models below offer only {\em relative} flow-cost valuations
(\eg{}, flow A is twice as costly as flow B); they do not operate
on absolute costs. These relative costs must be reconciled with the
blended prices used to derive customer valuations. We describe methods
for reconciling these values in Section~\ref{sec:mapping}. Each cost
model has a generic tuning parameter, denoted as $\theta$, which we use
in the evaluation.

\para{Linear function of distance.} The most straightforward way to model
ISP's costs as a function of distance is to assume cost increases
linearly with distance. Although, in some cases, this model does not hold
(\eg{}, crossing a mountain range is more expensive than crossing a
region with flat terrain), we often observe that ISPs charge linearly in
the distance of communication~\cite{www-chunghwa-pricelist,
www-bsnl-pricelist}.  As we model cost as linear function of distance, we
set cost $c_i = \gamma d_i + \beta$, where $\gamma$ is a scaling
coefficient that translates from relative to real costs, $\beta$ is a
base cost (\ie{}, the fixed cost that the ISP incurs for communicating
over any distance), and $d_i$ is the geographical distance between the
source and destination served by an ISP. We describe how we determine
$\gamma$ in Section~\ref{sec:mapping}. We model the base cost $\beta$ as
a fraction of the maximum cost without the base component.  More
formally: $\beta = \theta \max_{j\in 1\cdots n}\gamma d_j$, where
$\theta$ in this cost model is a relative base cost fraction, and $n$
is number of flows with different cost.  For example, given distances 1,
10, and 100 miles, $\gamma=\$1/mile$, and $\theta=0.1$, the resulting base
cost $\beta$ is \$10, and thus flow costs are \$11, \$20, and \$110. In
the evaluation, we vary $\theta$ to observe the effects of different base
costs. 
For example, low $\theta$ values (low base cost) here represent a
case where link distance is the largest contributor to the total cost.

\para{Concave function of distance.} We are also aware of ISPs that price
transit as a concave function of distance, which suggests that costs
likely follow the same pattern. For this scenario we model the ISP's cost
as $c_i = \gamma( a\log_b d_i + c) + \beta$. By fitting real pricing data
that follows concave pricing
patterns~\cite{www-itu-handbook,www-ntt-pricelist}, we find $a\approx
0.5$, $b\approx 6$, and $c\approx 1$ for normalized prices and distance.
As in the case of linear cost, we set the base offset cost $\beta =
\theta \max_{j\in 1\cdots n} \gamma(a\log_b d_i + c)$. We use $\theta$ in
the evaluation to change link distance contribution to the total cost.
\begin{figure}[t!]
  \centering
  \includegraphics[width=0.7\linewidth]{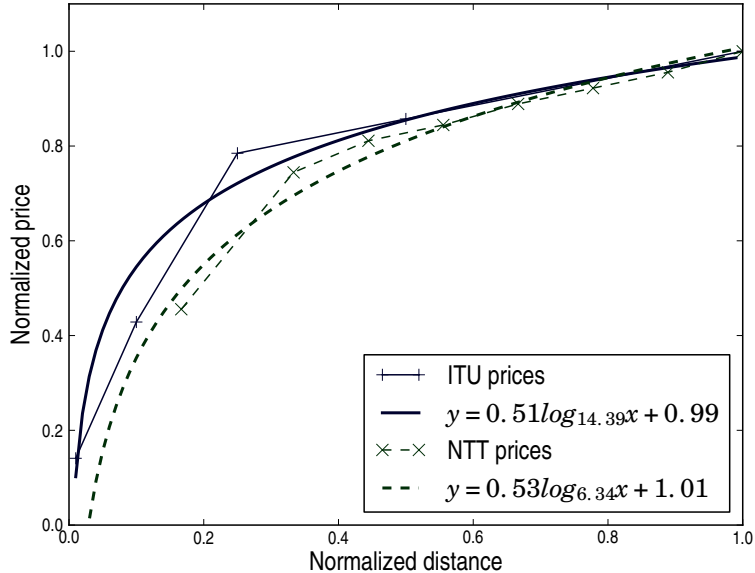}
  \caption{Using price data from ITU~\cite{www-itu-handbook}
  and NTT~\cite{www-ntt-pricelist} to fit concave distance to cost mapping curve.}
  \label{fig:log-price}
\end{figure}
Figure~\ref{fig:log-price} shows a concave curve fitting to two price
data sets, resulting in $a\approx 0.5$, $b\approx 6$, and $c\approx 1$
for normalized prices and distance. As in the case of linear cost, we set the
base offset cost $\beta = \theta \max_{j\in 1\cdots n} \gamma(a\log_b d_i
+ c)$. We use $\theta$ in the evaluation to change link distance
contribution to the total cost.

\para{Function of destination region.} Both private communication with
network operators and publicly available data suggest that ISPs can also
charge for traffic at rates depending on the region where traffic is
destined~\cite{www-etisalat-pricelist, www-ore-pricelist,
www-telegeography-prices}. For example, an ISP might have less expensive
capacity in the metropolitan area than in the region, less expensive
capacity in the region than in the nation, and less expensive capacity in
the nation than across continental boundaries. We divide flows into three
categories: {\em metropolitan}, {\em national}, and {\em international}.
We map the flows into these categories by using data from the
GeoIP~\cite{www-geoip} database: flows that originate and terminate in
the same city are classified as metro, and flows that start and end in
the same country are classified as national; all other flows are
classified as international. For EU ISP we only have distances between
traffic entry and exit points, thus we classify flows traveling less than
10 miles as metro, flows that travel less than 100 miles as national. We
set the costs as follows: $c_{metro}=\gamma$, $c_{nation}=\gamma
2^\theta$, and $c_{int}=\gamma 3^\theta$. This form allows us to test
scenarios when there is no cost difference between regions ($\theta=0$),
the cost differences are linear ($\theta=1$), and costs are different by
magnitudes ($\theta > 1$).

\para{Function of destination type.} ISPs often offer discounts for the
traffic destined to their customers (``on net'' traffic), while charging
higher rates for traffic destined for their peers (``off net'' traffic).
These offerings are motivated by the fact that ISPs do recover some of
their transport cost for the traffic sent to other customers. In our
evaluation, we model this cost difference by setting the cost of the
traffic to peers to be twice as costly than traffic to other customers.
The logic behind such a model is that when an ISP sends the traffic
between two customers, it gets paid twice by both customers, but when an
ISP sends traffic between a customer and a peer it is only paid by the
customer. The parameter $\theta$ indicates a fraction of traffic at each
distance that is destined to clients, as opposed to traffic that is
destined to peers and providers.

\section{Evaluating Tiered Pricing}
\label{sec:eval}

\begin{figure}
\centering
\includegraphics[width=0.7\linewidth]{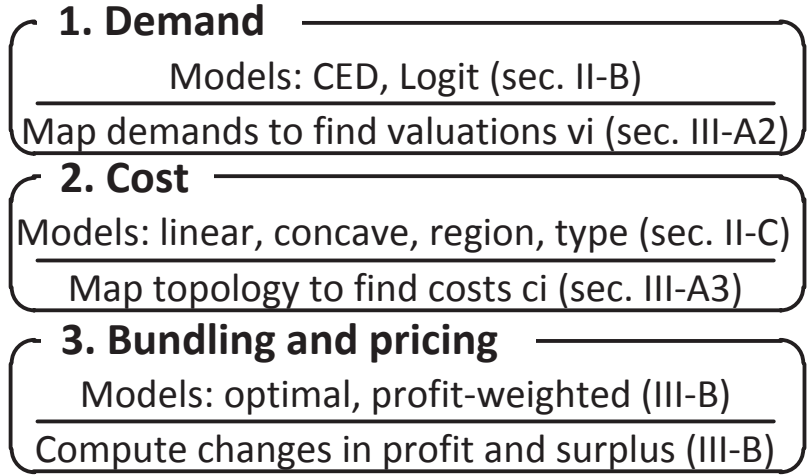}
\caption{We evaluate the effect of tiered pricing on Internet transit by
separately modeling the demand and cost of traffic. At each step we use real-world data to derive
unknown parameters.}
\label{fig:transit-model}
\end{figure}

In this section, we evaluate the efficiency of destination-based tiered
pricing using the model presented in Section~\ref{sec:model} and real
topology and demand data from large  networks.  Our goal is to understand
how the consumer surplus and the profit that an ISP extracts from
offering tiered-pricing depends on the number of tiers and the network
topology and traffic demand. One of the major challenges we face is that we cannot know some aspects
of the cost and demand models, or even which model to use. We use the ISP
data to derive model parameters, such as valuation or cost, and evaluate
the profit of each strategy across models and input parameters.
Figure~\ref{fig:transit-model} presents an overview of our approach.

Our evaluation yields several important results. First, we show that an
ISP needs only 3--4 bundles to capture 90--95\% of the profit provided by
an infinite number of bundles, if it bundles the traffic appropriately.
Second, choosing a bundling strategy that considers both flow demand and
cost is almost as effective as an exhaustive search for the best
combination of bundles. Finally, we observe that the topology and traffic
of a network influences its bundling strategies: networks with higher
coefficient of variation of demand need more bundles to extract maximum
profit.

\vspace{-3mm}
\subsection{Mapping Data to Models}\label{sec:mapping}
\label{sec:eval1}

Because we do not know the parameters that we need to compute the
profit-maximizing prices we must derive them.  We first describe the data
then we show how to apply the demand models to the real traffic demands
to compute the valuation coefficients $v_i$ for each flow $i$. Finally,
we derive the ISP's cost for servicing the flows by applying cost models
to the observed flow distances.

\begin{table}
\small
\centering
\begin{tabular}{|l||c|c|c|c|c|}
\hline
\multirow{2}{*}{\textbf{Data set}} 	& \multirow{2}{*}{\textbf{Date}} &
\multicolumn{2}{c|}{\textbf{Distance (miles)}} & \multicolumn{2}{c|}{\textbf{Traffic (Gbps)}}\\
\cline{3-6} &  & \textbf{w-avg} & \textbf{CV} & \textbf{Aggregate} & \textbf{CV}\\
\hline
EU ISP			& 11/12/09     & 54 & 0.70 & 37 & 1.71 \\
CDN			& 12/02/09     & 1988 & 0.59 & 96 & 2.28  \\
Internet 2		& 12/02/09     & 660 & 0.54 & 4 & 4.53 \\
\hline
\end{tabular}
\caption{Data sets used in our evaluation. The columns represent: the
network, data capture date, demand-weighted average of flow distances,
coefficient of variation (CV) of flow distances, aggregate traffic per
second, and CV of demand of different flows.}
\label{tab:datasets}
\end{table}


\subsubsection{Data Sources}
\label{sec:data}

We use demand and topology data from three networks: a European ISP
serving thousands of business customers (EU ISP), one of the largest CDN
providers in the world (CDN), and a major research network in United
States (Internet 2). The data consists of sampled NetFlow records from
core routers in each network for 24 hours.  Table~\ref{tab:datasets}
presents more details about the data sets.

To drive our model, we must compute the traffic volume (which captures
consumer demand) and the distance between the source and destination of
each flow (which captures the relative cost of transit).  To do so, we
extract the source and destination IP, as well as the traffic level, from
each NetFlow record. We obtain the demand for each flow by aggregating
all records of the flow, while ensuring that we do not double-count
record duplicated on different routers.

To compute distances that reflect the ISP's cost of sending traffic, we
use the following heuristics. For the EU ISP, the distance that each flow
travels in the ISP's network is the geographical distance between the
flow's entry and exit points, whose identity and location is known. For
the CDN, we use the GeoIP database~\cite{www-geoip} to estimate the
distance to the destination. Although this may not reflect the real
distance that a packet travels (because part of the path may be covered
by another ISP), we assume that it is still reflective of the cost
incurred by the CDN.  Finally, for Internet2 we use the router interface
information to identify the links the flow has traversed. The distance
each flow traverses is the sum of the links in the path, where the link
length is the geographical distance between the neighboring routers.

\subsubsection{Discovering valuation coefficients}

The valuation coefficient $v_i$ indicates the valuation of flow $i$. We
need the valuation coefficeint $v_i$ to capture how the demand for flow
$i$ varies with price (Equations~\ref{eqn:demand-ced}
and~\ref{eqn:demand-logit}) and thus affect the ISP profit. To find $v_i$
for each flow, we assume that ISPs charge the same blended price $P_0$
for each flow and map the observed traffic demand from the data to each
demand model. 

\para{CED valuation coefficient.} From Equation~\ref{eqn:demand-ced} we
obtain:

\vspace{-2mm}
\begin{equation}
  \textstyle v_i = P_0q_i^{1/\alpha},
\end{equation}

\noindent where $q_i$ is observed demand on flow $i$ and $\alpha$ is the
sensitivity coefficient that we vary in the evaluation.

\para{Logit demand valuation coefficient.} Dividing both left and right
sides of market share expression (Equation~(\ref{eqn:marketshare}) by the
corresponding sides of $s_0$ (Equation~(\ref{eqn:share-zero})), we can
find valuation of flow $i$:

\vspace{-2mm}
\begin{equation}
  \textstyle v_i = \frac{\log s_i - \log s_0}{\alpha} + P_0,
\end{equation}

\noindent where $s_i$ is the market share of flow $i$. We vary $s_0$ in the
evaluation and compute the remaining market shares from observed traffic
as $s_i = \frac{q_i(1-s_0)}{\sum q_i}$.

\subsubsection{Discovering costs}
\label{sec:reconsile}

To estimate the ISP profit, we must know the cost that an ISP incurs to
service each flow. However, the data provides information only about the
distance each flow traverses, which reflects only the relative cost
(\eg{}, flow A is twice as costly as flow B), rather than an absolute
cost value. To normalize the cost of carrying traffic to the same units
as the price for the flow, we introduce a scaling parameter $\gamma$,
where $c_i = \gamma f(d_i)$ and $d_i$ is the distance covered by $i$ (see
Section~\ref{sec:cost-model}). For each demand model, we compute the
scaling parameter by assuming (as in computing valuation coefficients)
that ISPs are rational and profit maximizing, and charge the same
original price $P_0$ for each flow.

\para{CED.} Differentiating profit Equation~(\ref{eqn:profit-ced}) in
terms of price and then substituting $c_i$ with $\gamma f(d_i)$, we find
cost normalization coefficient:

\vspace{-3mm}
\begin{equation}
  \gamma = \frac{P_0(\alpha-1)\sum_{i=1}^n v_i^\alpha}{\alpha\sum_{i=1}^n
  f(d_i)v_i^\alpha}
\end{equation}

\para{Logit demand.} Like in CED case, differentiating profit
Equation~\ref{eqn:logit-profit} and substituting $c_i$ for $\gamma
f(d_i)$, we can express $\gamma$:

\vspace{-3mm}
\begin{equation}
  \gamma = \frac{\sum_{i=1}^n\left(e^{\alpha(v_i-P_0)}\left(\alpha P_0 - 1
  - \sum_{j=1}^n e^{\alpha(v_j-P_0)}\right)\right)}{\alpha \sum_{i=1}^n
  f(d_i)e^{\alpha(v_i-P_0)} }
\end{equation}

\vspace{-3mm}

\subsection{Effects of Tiered Pricing}
\label{sec:bundling}

ISPs must judiciously choose how they bundle traffic flows into tiers. As
shown in Section~\ref{sec:services}, today's ISPs often offer at most two
or three bundles with different prices. We define six bundling strategies
that classify and group traffic flows according to their cost, demand, or
potential profit to the ISP. We then evaluate them and show that,
assuming the right bundling strategy is used, ISPs typically need only a
few bundles to collect near-optimal profit.

\begin{figure*}
  \subfigure[European
  ISP.]{\includegraphics[width=0.33\linewidth]{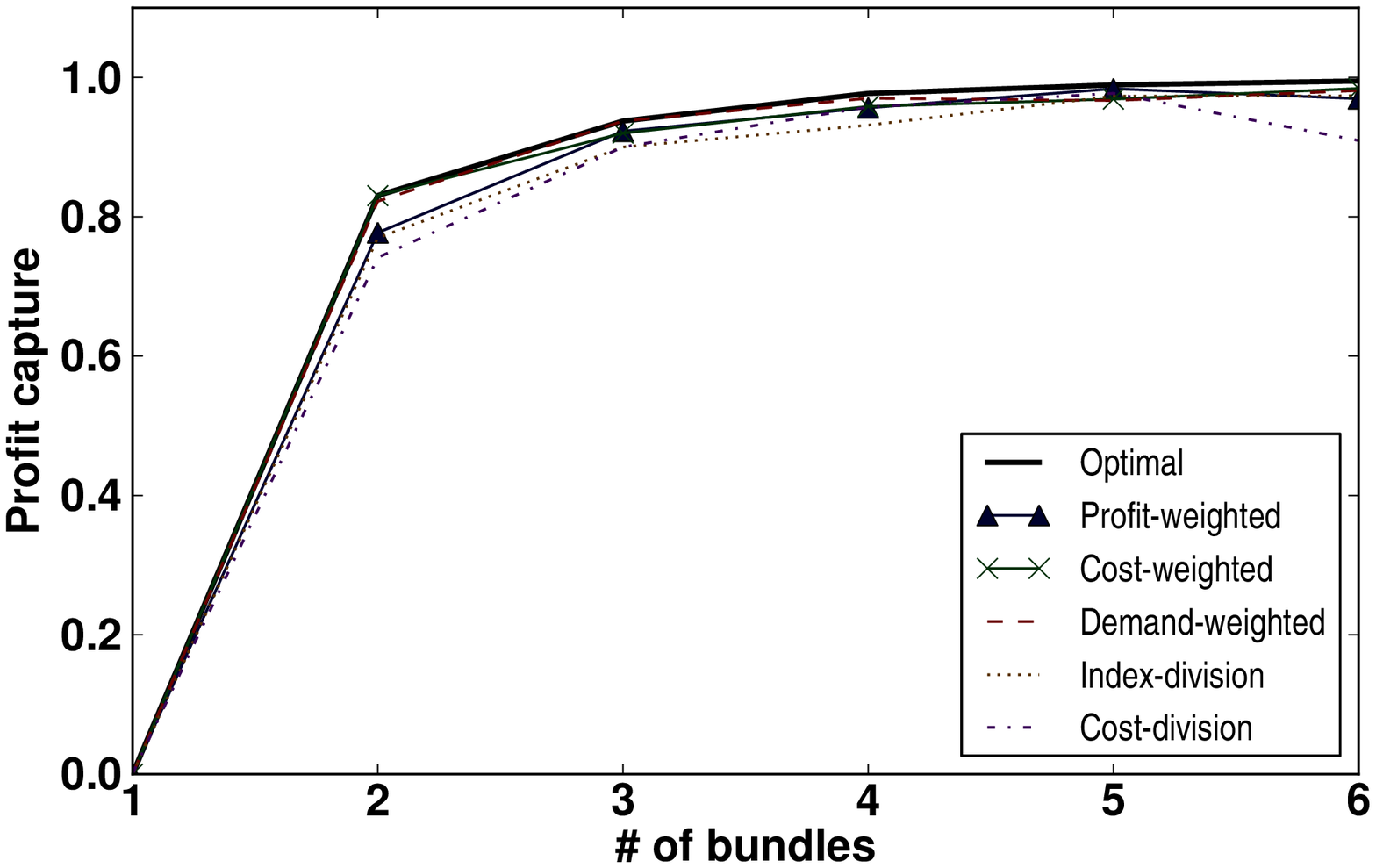}}
  \subfigure[Internet2.]{\includegraphics[width=0.33\linewidth]{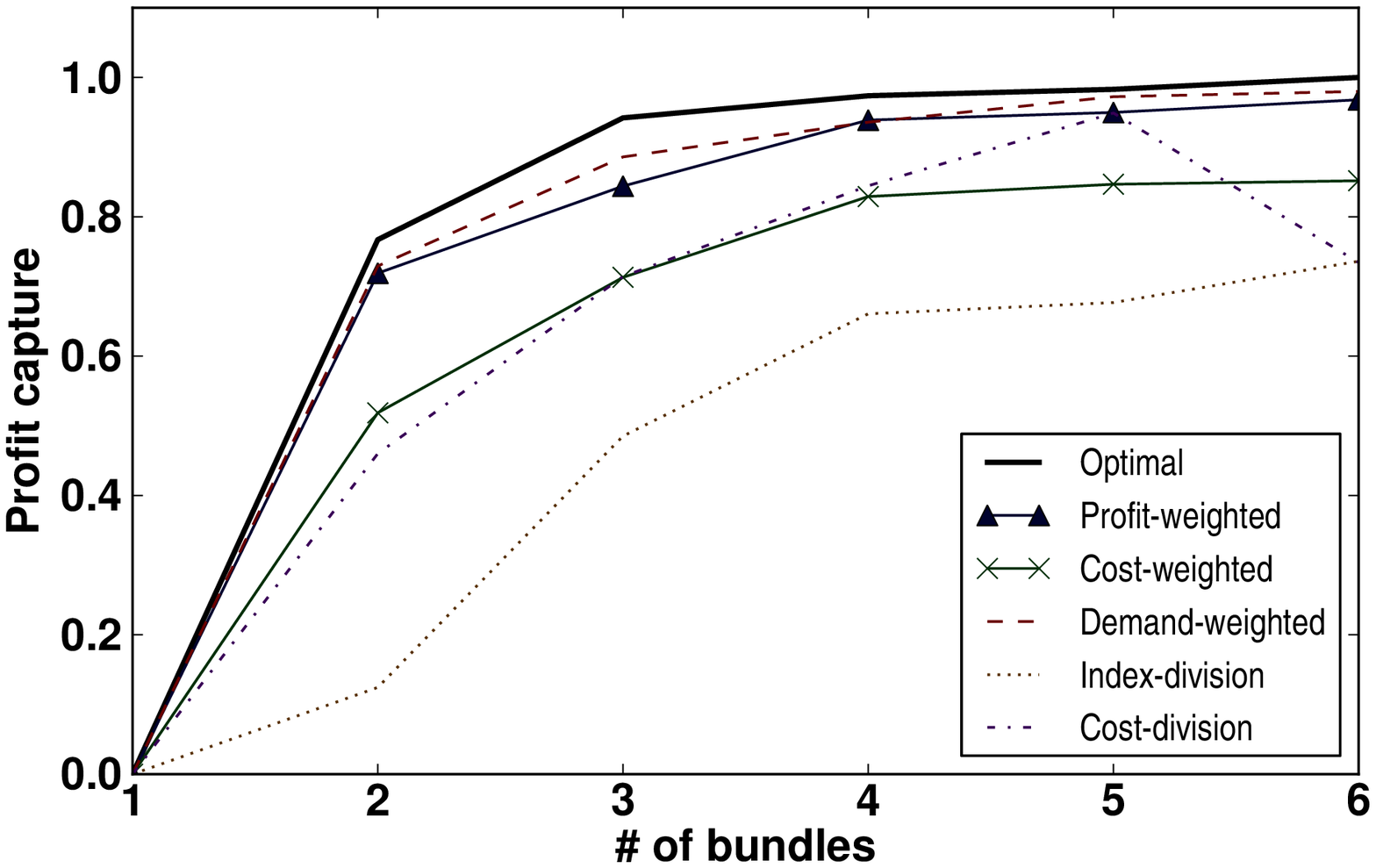}}
  \subfigure[International
  CDN.]{\includegraphics[width=0.33\linewidth]{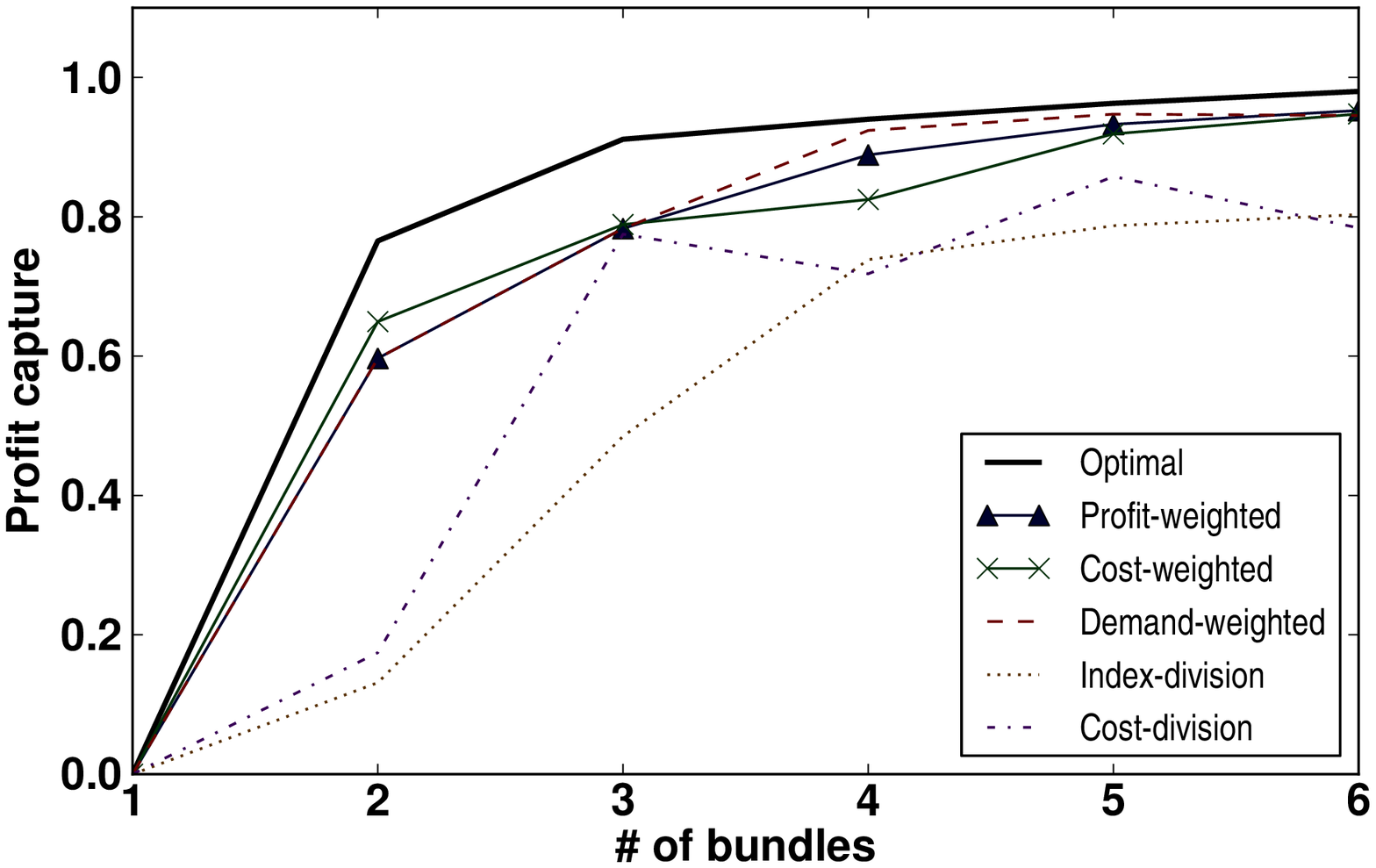}}
  \caption{Profit capture for different bundling strategies in constant
  elasticity demand.}
  \label{fig:ced-division-types}
\end{figure*}

\begin{figure*}
  \subfigure[European ISP.]{\includegraphics[width=0.33\linewidth]{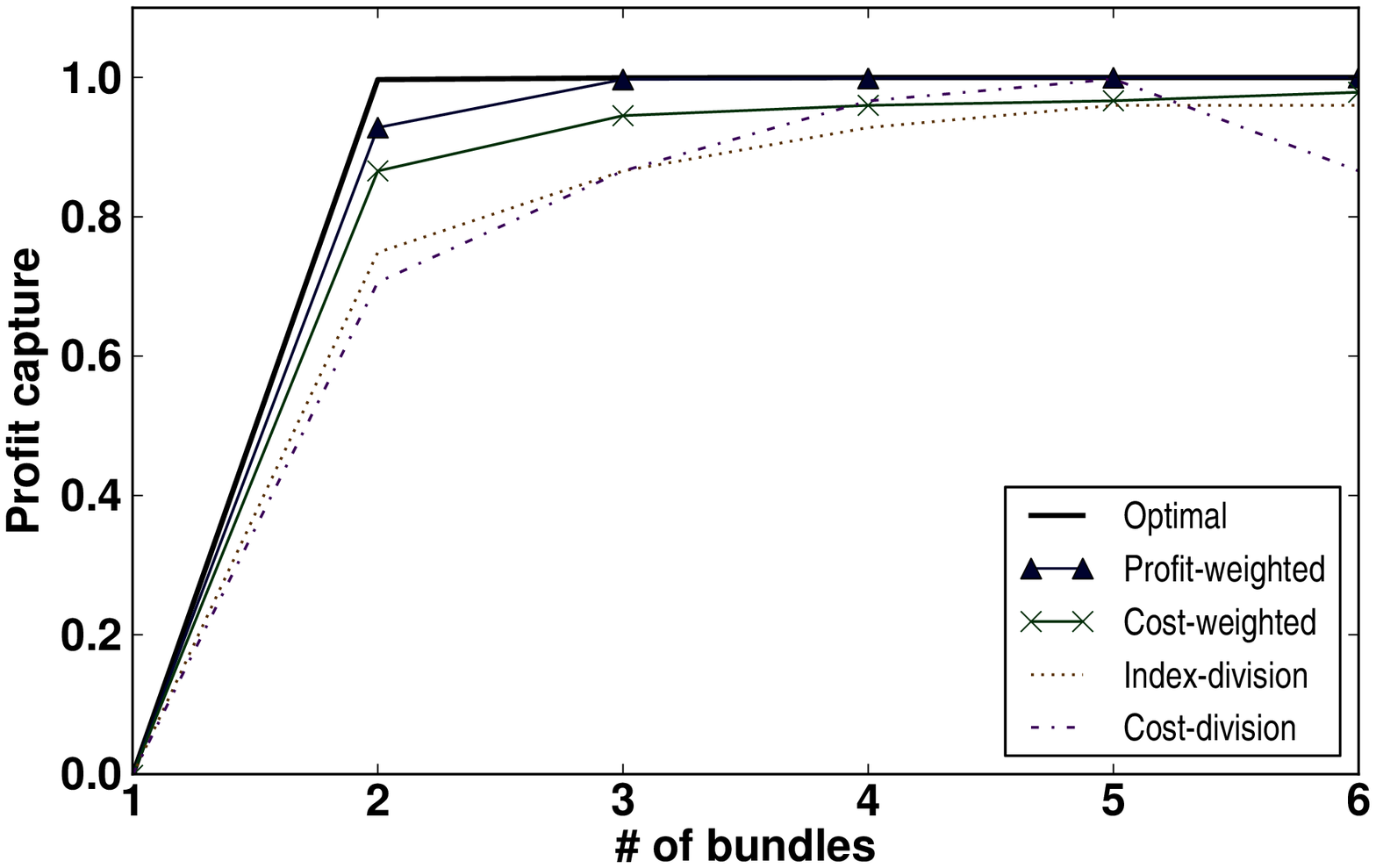}}
  \subfigure[Internet2.]{\includegraphics[width=0.33\linewidth]{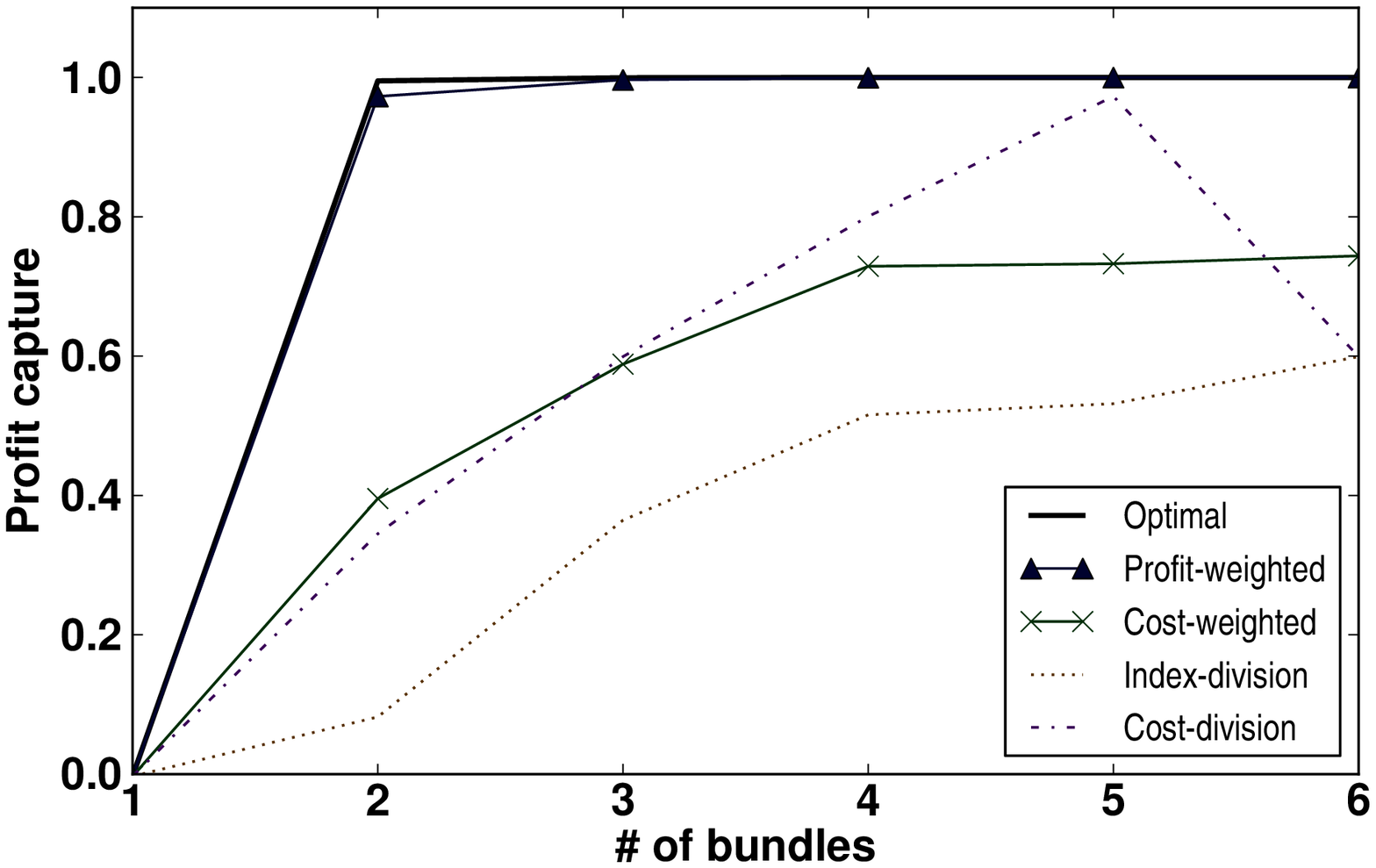}}
  \subfigure[International CDN.]{\includegraphics[width=0.33\linewidth]{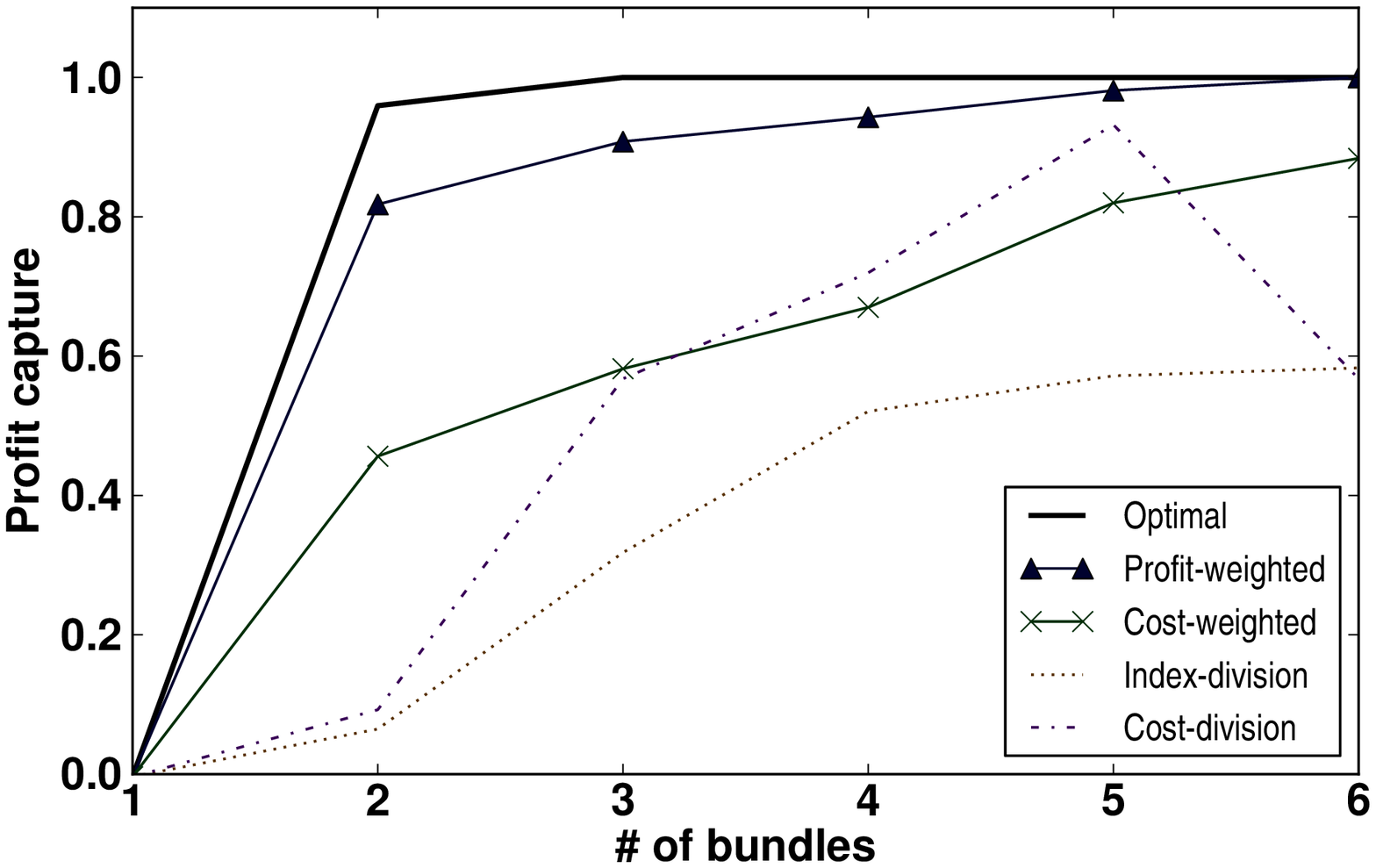}}
  \caption{Profit capture for different bundling strategies in logit demand.}
  \label{fig:logit-division-types}
\end{figure*}

\subsubsection{Bundling strategies}
\label{sec:bundling-strategies}

\para{Optimal.} We exhaustively search all possible combinations of
bundles to find the one that yields the most profit. This approach gives
optimal results and also serves as our {\em baseline} against which we
compare other strategies. Computing the optimal bundling is
computationally expensive: for example, there is more than a billion ways
to divide one hundred traffic flows into six pricing bundles. Presented
below, all of the other bundling strategies employ heuristics to make
bundling computationally tractable.

\para{Demand-weighted.} In this strategy, we use an algorithm inspired
by token buckets to group traffic flows to bundles. First, we set the
overall token budget as the sum of the original demand of all flows:
$T=\sum_i q_i$. Then, for each bundle $j$ we assign the same token
budget $t_j=T/B$, where $B$ is the number of bundles we want to create.
We sort the flows in decreasing order of their demand and traverse them
one-by-one. When traversing flow $i$, we assign it to the first bundle
$j$ that either has no flows assigned to it or has a budget $t_j>0$. We
reduce the budget of that bundle by $q_i$. If the resulting budget
$t_j<0$, we set $t_{j+1}=t_{j+1} + t_j$. After traversing all the flows,
the token budget of every bundle will be zero, and each flow will be
assigned to a bundle. The algorithm leads to separate bundles for high
demand flows and shared bundles for low demand flows. For example, if we
need to divide four flows with demands 30, 10, 10, and 10 into two
bundles, the algorithm will place the first flow in the first bundle,
and the other three flows in the second bundle.

\para{Cost-weighted.} We use the same approach as in demand-weighted
bundling, but we set the token budget to $T=\sum_i 1/c_i$. When placing
a flow in a bundle we remove a number of tokens equal to the inverse of
its cost.  This approach creates separate bundles for local flows and
shared bundles for flows traversing longer distances. The current ISP
practices of offering {\em regional pricing} and {\em backplane peering}
maps closely to using just two or three bundles arranged using this
cost-weighted strategy.

\para{Profit-weighted.} The bundling algorithms described above consider
cost and demand separately. To account for cost and demand together, we
estimate {\em potential profit} each flow could bring. We use the
potential profit metric to apply the same weighting algorithm as in cost
and demand-weighted bundling. In case of constant elasticity demand, we
derive potential profit of each flow $i$:

\vspace{-4mm}
\begin{equation}
  \pi_i = \frac{v_i^\alpha}{\alpha}
  \left(\frac{\alpha c_i}{\alpha-1}\right)^{1-\alpha}
\end{equation}

For the logit demand, substituting $p_i$ in
Equation~\ref{eqn:logitprofit} yields:

\begin{equation}
  \pi_i = Ks_i(p_i-c_i) = \frac{Ks_i}{\alpha s_0}\propto q_i
\end{equation}


\para{Cost division.} We find the most expensive flow and divide
the cost into ranges according to that value.  For example, if we want to
introduce two bundles and the most expensive flow costs
\$10/Mbps/month to reach, we assign flows that cost \$0--\$4.99 to the
first bundle and flows that cost \$5--\$10 to the second bundle.

\para{Index division.} Index-division bundling is similar to cost
division bundling, except that we rank flows according to their cost and
use the rank, rather than the cost, to perform the division into
bundles.

\subsubsection{The effects of different bundling strategies}
\label{sec:effects1}

To evaluate the bundling strategies described above, we compute the
profit-maximizing prices and measure the resulting pricing outcome in
terms of {\em profit capture}. Profit capture indicates what fraction of
the maximum possible profit---the profit attained using an infinite
number of bundles---the strategy captures. For example, if the maximum
attainable profit is 30\% higher than the original profit, while the
profit from using two bundles is 15\% higher than the original profit,
the profit capture with two bundles attains 0.5 of profit capture.
Formally, profit capture is $(\pi_{\mathsf{new}} -
\pi_{\mathsf{original}})/(\pi_{\mathsf{max}}-\pi_{\mathsf{original}})$.

Figures~\ref{fig:ced-division-types} and~\ref{fig:logit-division-types}
show the profit capture for different bundling strategies, across the
three data sets, while varying the number of bundles. For the results
shown here, we use both the constant elasticity and the logit demand
models and the linear cost model. We set the price sensitivity $\alpha$
to $1.1$, the original, blended rate $P_0$ to $\$20$, the cost tuning
parameter $\theta$ to $0.2$, and the original market fraction that sends no
traffic $s_0$ to $0.2$.  We explore the effect of varying these
parameters in Section~\ref{sec:sensitivity}.

\para{Optimal versus heuristics-based bundling.} With an appropriate
bundling strategy, the ISP attains maximum profit with just 3--4
bundles.  As expected, the optimal flow bundling strategy captures the
most profit for a given number of bundles. We observe that the EU ISP
captures more profit with two bundles than other networks. We attribute
this effect to the low coefficient of variation (CV) of demand to
different destinations, which limits the benefits of having more pricing
bundles. We also discover that, given fixed demand, a high CV of
distance (cost) leads to higher absolute profits. With only minor exceptions,
the profit-weighted bundling heuristic is almost as good as the the
optimal bundling, followed by the cost-weighted bundling
heuristic. Deeper analysis, beyond the scope of this work, could show
what specific input data conditions cause the profit-weighted flow
bundling heuristic to produce bundlings superior to the cost-weighted
heuristic.

\para{Logit profit capture.} Maximum profit capture occurs more quickly
in the logit model because (1)~the total demand (including $s_0$ option)
is constant, and (2)~the model is sensitive to differences in
valuation of different flows. When there is a flow with a significantly
higher difference between valuation and cost ($v_i-c_i$), it absorbs most
of the demand. In this model, with just two pricing tiers, local and
non-local traffic are separated into distinct bundles that closely
represent the {\em backplane peering} and {\em regional pricing} for
local area service models.

\begin{figure}[h!]
  \centering
  \subfigure[Constant elasticity
  demand.]{\includegraphics[width=0.7\linewidth]{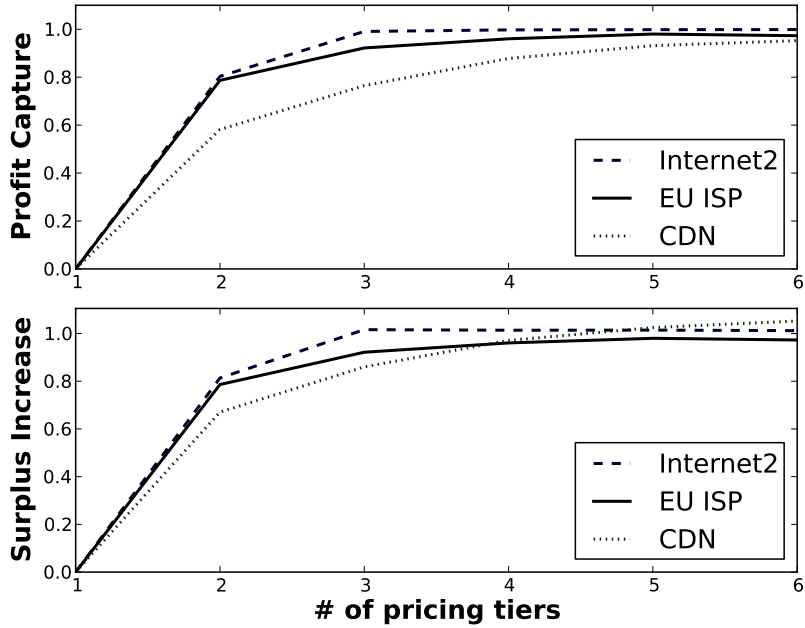}}
  \subfigure[Logit.]{\includegraphics[width=0.7\linewidth]{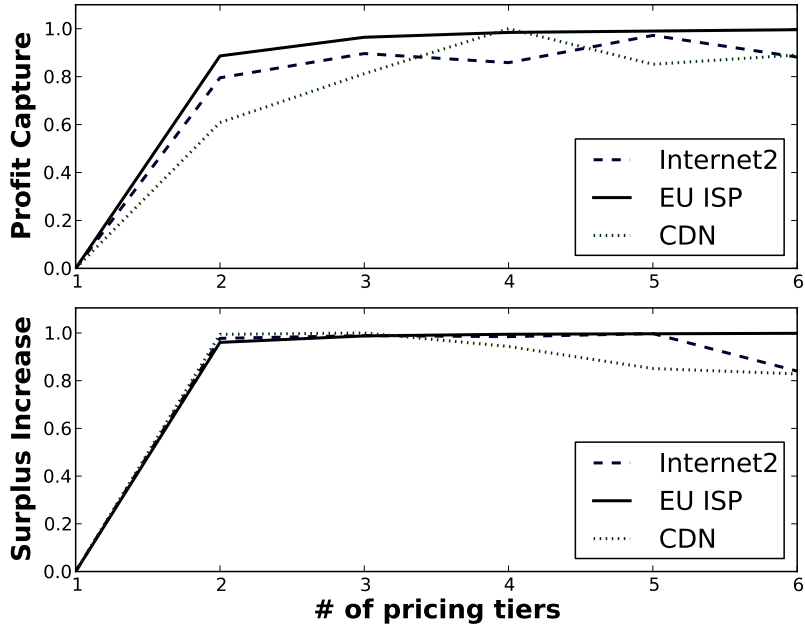}}
  \caption{ISP profit and consumer surplus change as ISP employ an
  increasing number of pricing tiers. Concave cost model, $\theta=0.2$,
  $\alpha=1.1$, $s_0=0.2$.}
  \label{fig:profit-capture}
\end{figure}

\subsubsection{Pricing effects on consumer surplus}

Figure~\ref{fig:profit-capture} contrasts consumer surplus change next to
ISP profit gains. The consumer surplus here is normalized to the surplus
gain the consumers get when ISPs maximize their profit with an infinite
number of pricing tiers. As expected, the surplus gain follows closely,
if not precisely, ISP profit gains.

\begin{figure}[h!]
  \centering
  \includegraphics[width=0.7\linewidth]{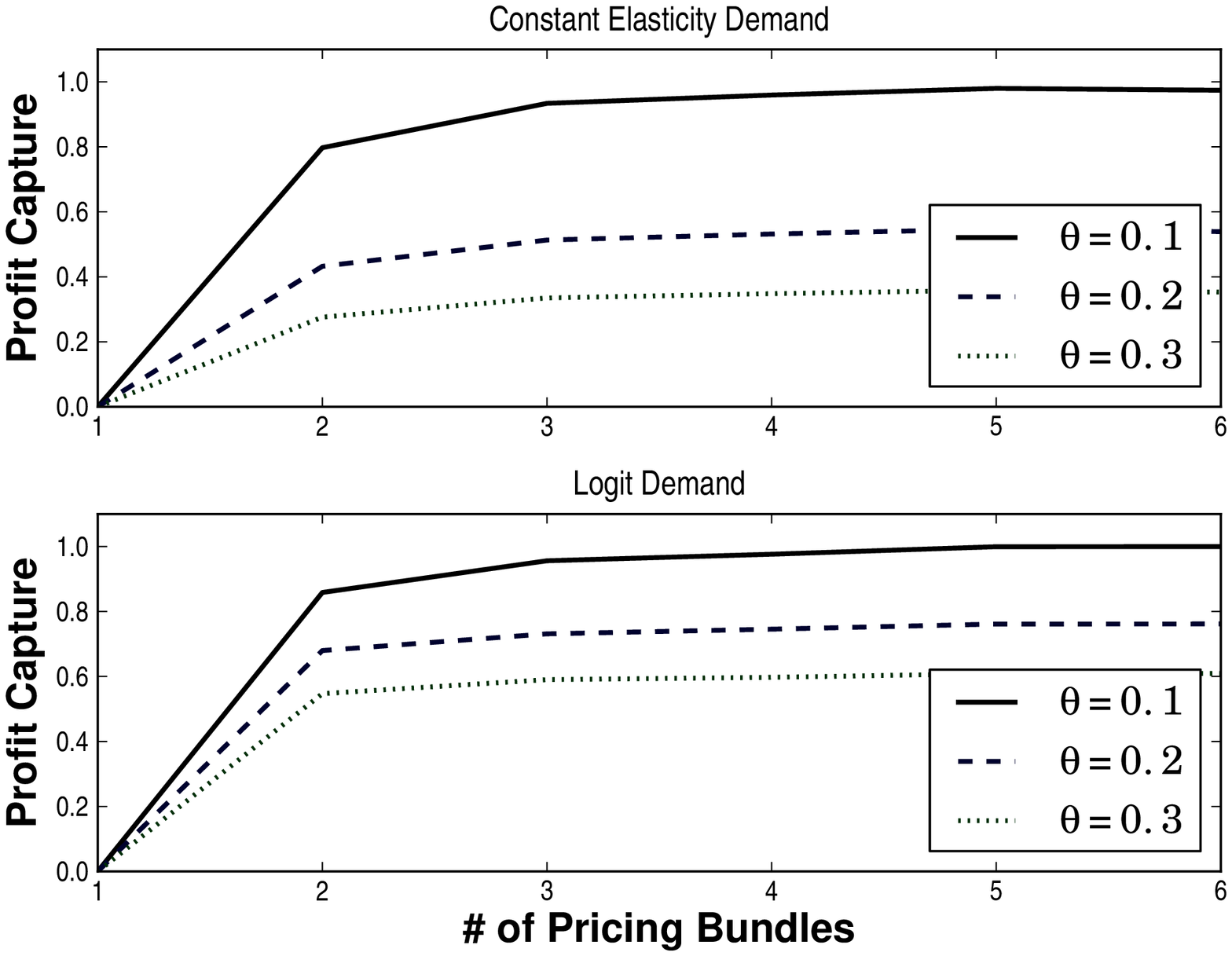}
  \caption{Profit increase in EU ISP network using linear cost model.}
  \label{fig:linear-cost}
\end{figure}

\begin{figure}[h!]
  \centering
  \includegraphics[width=0.7\linewidth]{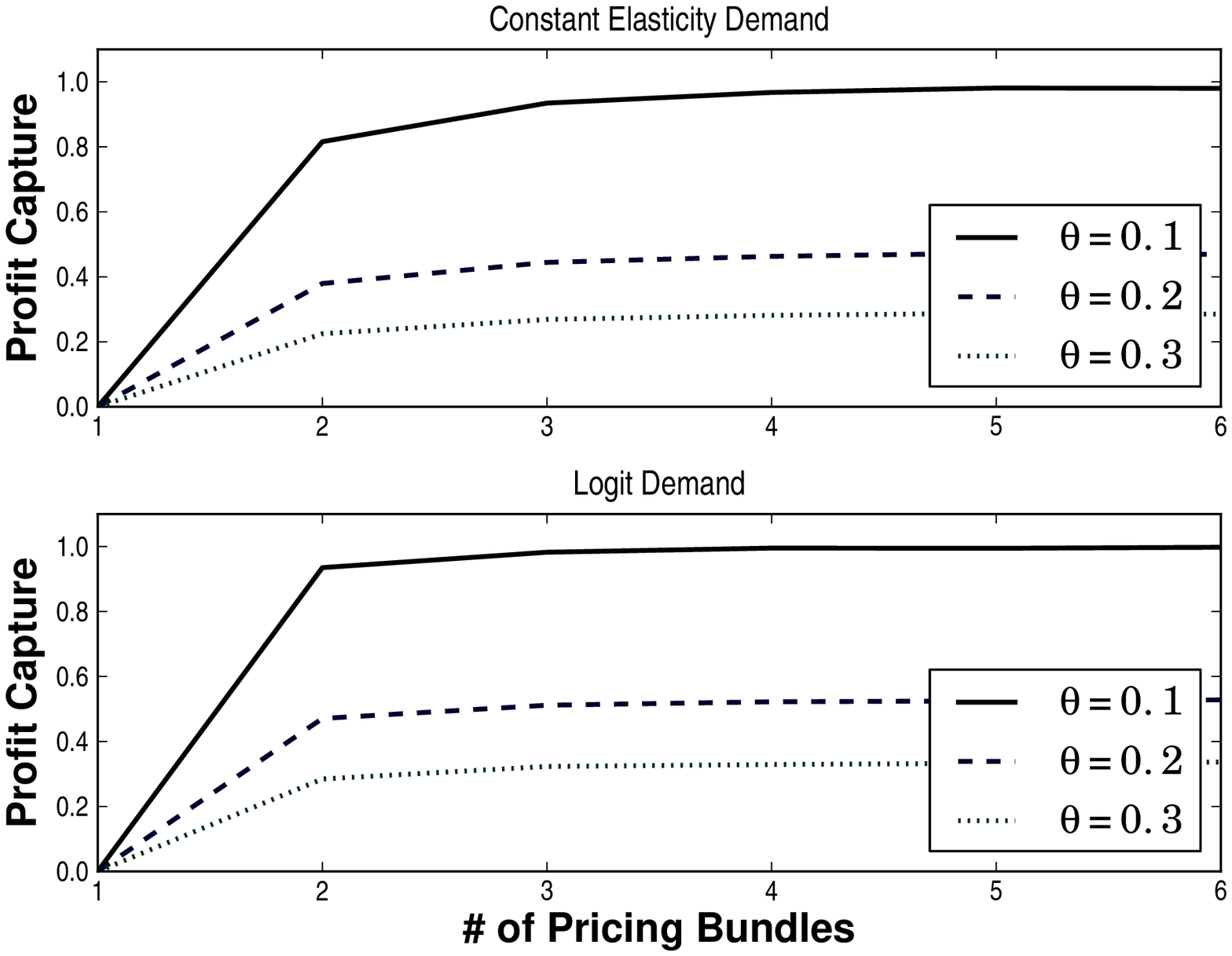}
  \caption{Profit increase in EU ISP network using concave cost model.}
  \label{fig:concave-cost}
\end{figure}

\subsection{Sensitivity Analysis}
\label{sec:sensitivity}

We explore the robustness of our results to cost models and input
parameter settings. As we vary an input parameter under test, other
parameters remain constant. Unless otherwise noted, we use
profit-weighted bundling, the EU ISP dataset, sensitivity $\alpha=1.1$,
the linear cost model with base cost $\theta=0.2$, blended rate
$P_0=\$20.0$, and, in the logit model, $s_0=0.2$ (the original market
fraction that sends no traffic).

\subsubsection{Effects of cost models}
\label{sec:cost-eval}

\begin{figure}[h!]
  \centering
  \includegraphics[width=0.7\linewidth]{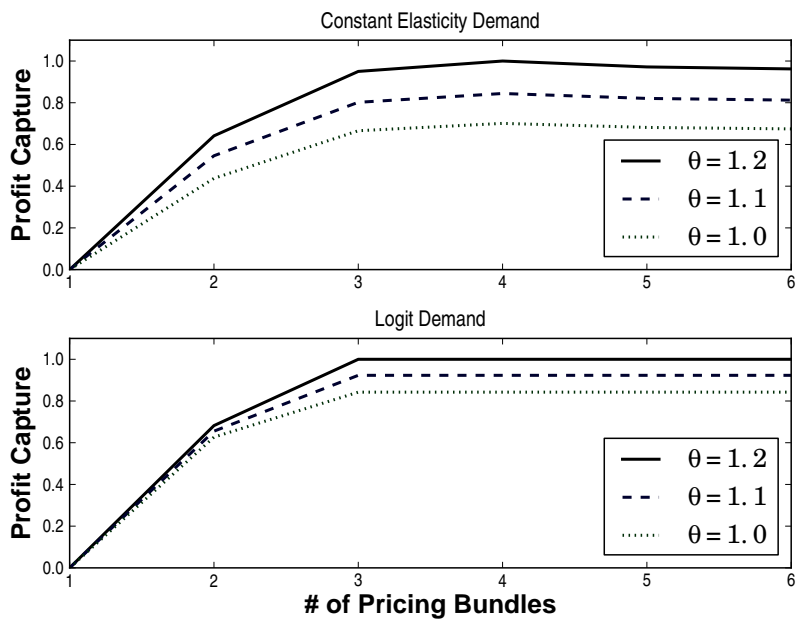}
  \caption{Profit increase in the EU ISP network using  regional cost model.}
  \label{fig:regional-cost}
\end{figure}

\begin{figure}[h!]
  \centering
  \includegraphics[width=0.7\linewidth]{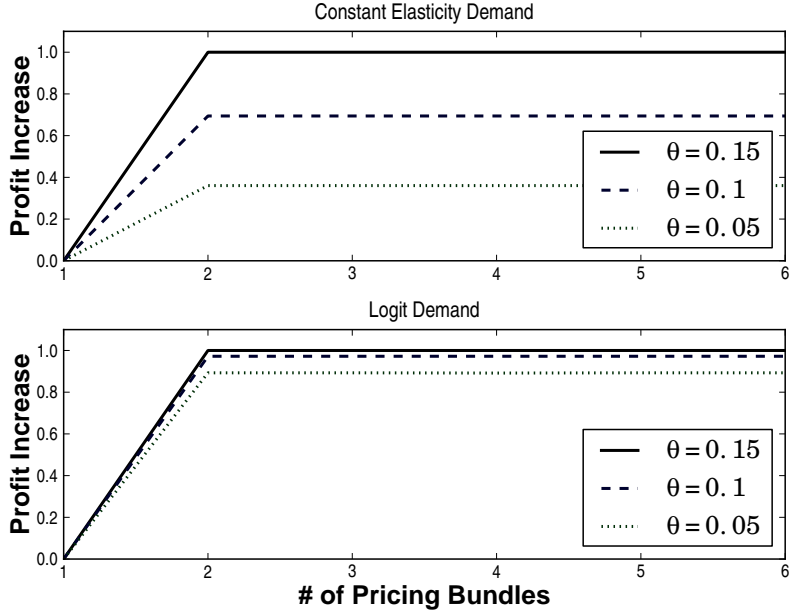}
  \caption{Profit increase in the EU ISP network using destination type cost model.}
  \label{fig:type-cost}
\end{figure}

We aim to see how cost models and settings within these models
qualitatively affect our results from the previous section. We show how
profit changes as we increase the number of bundles for different
settings of the cost model parameters ($\theta$), described in
Section~\ref{sec:cost-model}. We find that for different $\theta$
settings most of the attainable profit is still captured in 2-3 bundles.
Unlike in other sections, in
Figures~\ref{fig:linear-cost}--\ref{fig:type-cost},
we normalize the profit of all the plots in the
graphs to the highest observed profit. In other words, $\pi_{max}$ in
these figures is not the maximum profit of each plot, but the maximum
profit of the plot with highest profit in the figure. Normalizing
by the highest observed profit allows us to show how changing
the parameter $\theta$ affects the amount of profit that the ISP can
capture.

\para{Linear cost.} 
Figure~\ref{fig:linear-cost} shows profit increase in the EU ISP network
as we vary the number of bundles for different settings of $\theta$. As
expected, most of the profit is still attained with 2--3 pricing bundles.
We also observe that the increase in the base cost ($\theta$) causes
a decline in the maximum attainable profit. The reduction in maximum
attainable profit is expected, as increasing the base cost reduces the
coefficient of variation (CV) of the cost of different flows and thus reduces
the opportunities for variable pricing and profit capture. We can also
see, as shown in previous section, that the logit demand model attains
more profit than the constant elasticity demand model with the same
number of pricing bundles.

\para{Concave cost.} Figure~\ref{fig:concave-cost} shows the profit
increase as we vary the number of bundles for different
settings of $\theta$ for the concave cost model. The observations and results
are similar to the linear cost model, with one notable exception. The amount
of profit the ISP can capture decreases more quickly in the concave cost model than in
the linear cost model for the same change in the base-cost parameter $\theta$.
This is due to the lower CV of cost in the concave model than in the linear cost
model. In other words, applying the log function on distance (as described in
Section~\ref{sec:cost-model}) reduces the relative cost difference between
flows traveling to local and remote destinations. 

\para{Regional cost.} In the regional cost model, the parameter $\theta$ is
an exponent which adjusts the price difference between three different
regions: local, national, and international.
Figure~\ref{fig:regional-cost} shows the profit increase in the EU ISP
network as we vary number of bundles for different settings of $\theta$.
Higher $\theta$ values result in a higher CV of cost in different
regions which, in turn, in both demand models produces higher profit.
Using constant elasticity demand we observe a small dip in profit when
using five and six bundles, which recovers later with more bundles. Such
dips are expected when there are only a few traffic classes. For example, if traffic
had just two distinct cost classes, two judiciously selected bundles
could capture most of the profit. Adding a third bundle can reduce the
profit if that third bundle contains flows from both of the classes
(as may happen in a suboptimal bundling).

\para{Destination type-based cost.} Destination type-based cost model
emulates ``on-net'' and ``off-net'' types of traffic in an ISP network.
As described in Section~\ref{sec:cost-model}, we assume that ``on-net''
traffic costs less than ``off-net'' traffic. We vary $\theta$, which
represents a fraction of ``on-net'' traffic in each flow. The standard
profit-weighting algorithm does not work well with the destination
type-based cost model. The effect observed in the regional cost
model---where five bundles produce slightly lower profit then four
bundles---is 
more pronounced when we have just two distinct flow classes. One
heuristic that works reasonably well is as follows: we update the
profit-weighting heuristic to never group traffic from two different
classes into the same bundle.  Figure~\ref{fig:type-cost} shows how
profit increases with an increasing number of bundles. Since there are
two major classes of traffic (``on-'' and ``off-net''), most profit is
attained with two bundles for both demand models. In this cost model, as
in other cost models, the same change in CV of cost (induced by the
parameter $\theta$) causes a greater change in profit capture for
constant elasticity demand than for logit demand.

\subsubsection{Sensitivity to parameter settings}

The models we use rely on a set of parameters, such as price sensitivity
($\alpha$), price of the original bundle ($P_0$), and, in the logit
model, the share of the market that corresponds to deciding not to
purchase bandwidth ($s_0$). In this section, we analyze how sensitive
the model is to the choice of these parameters.

\begin{figure}[h!]
  \centering
  \includegraphics[width=0.7\linewidth]{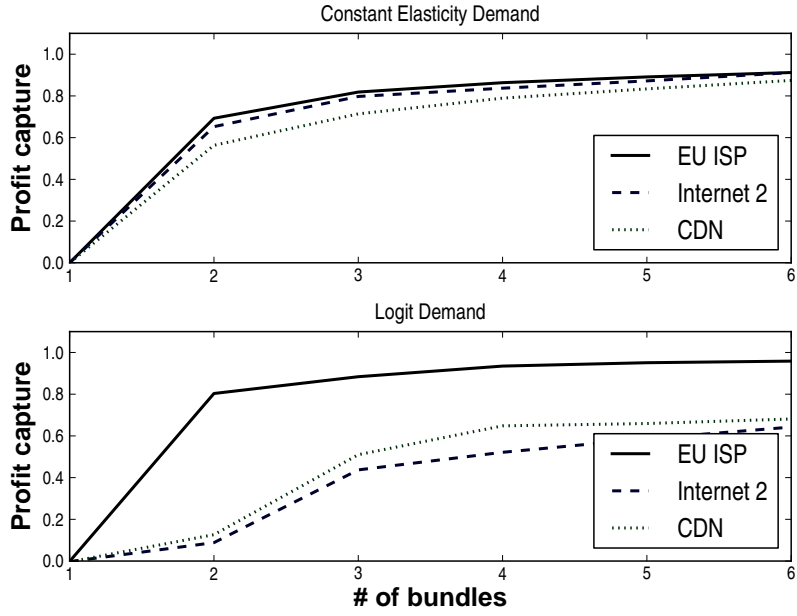}
  \caption{Minimum profit capture for a fixed number of bundles over a range of $\alpha$ between 1 and
  10.}
  \label{fig:alpha-sensitivity}
\end{figure}

\begin{figure}[h!]
  \centering
  \includegraphics[width=0.7\linewidth]{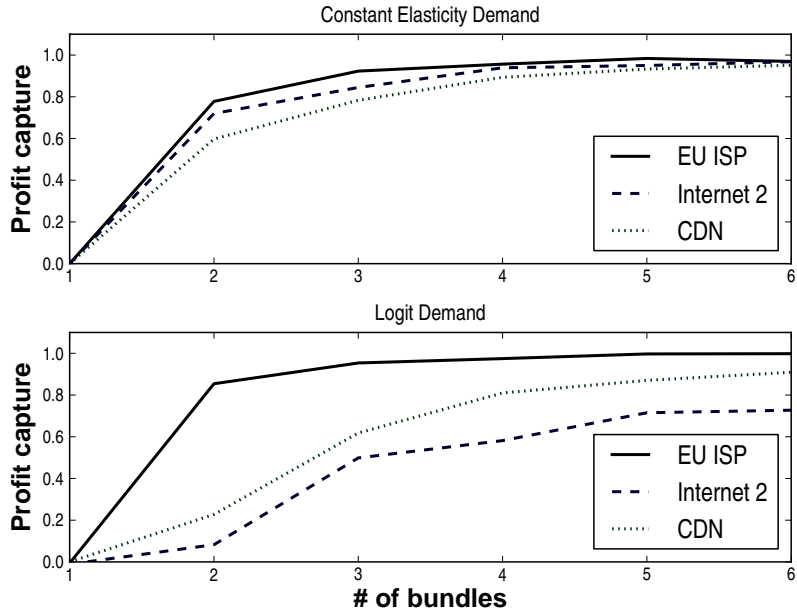}
  \caption{Minimum profit capture for a fixed number of bundles over a range of
  starting prices $P_0\in[5,30]$.}
  \label{fig:pzero-sensitivity}
\end{figure}

\begin{figure}[h!]
  \centering
  \includegraphics[width=0.7\linewidth]{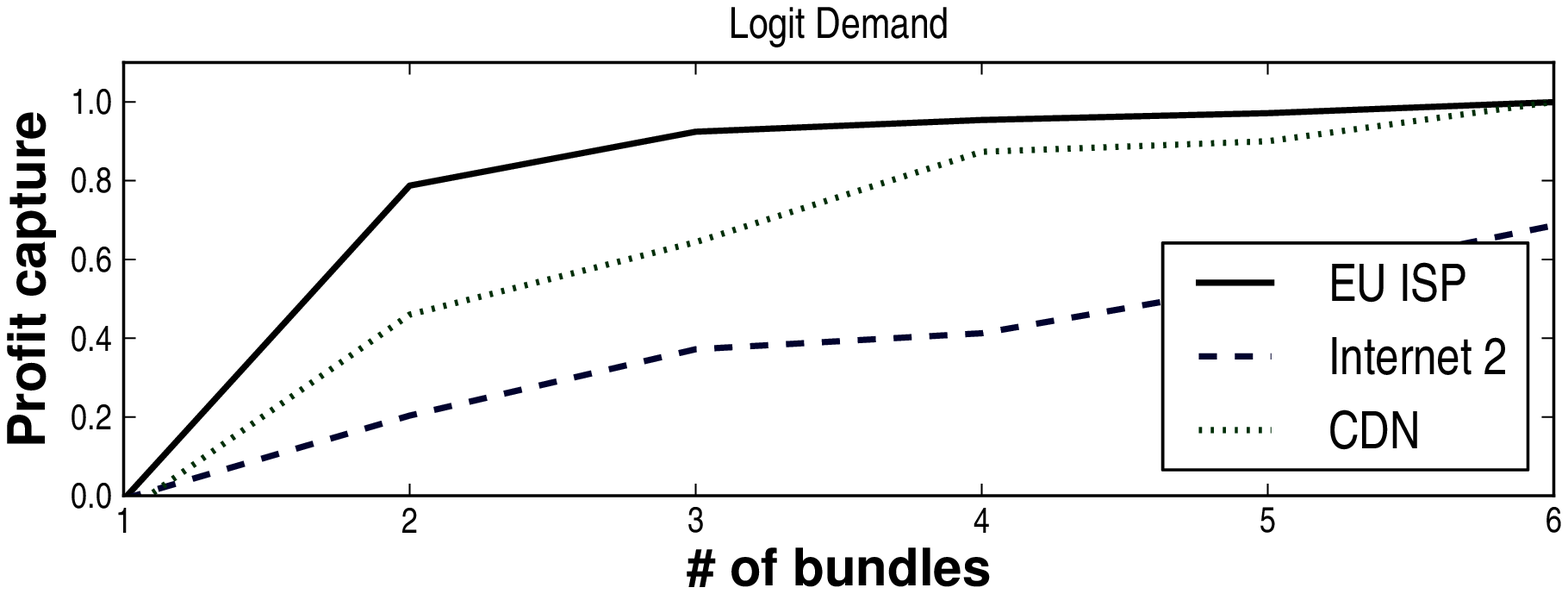}
  \caption{Maximum profit capture for a fixed number of bundles over a range 
  of fractions of users who decide not to participate in the market
  $s_0\in(0,1)$.}
  \label{fig:szero-sensitivity}
\end{figure}

Figures~\ref{fig:alpha-sensitivity}--\ref{fig:szero-sensitivity}
show how profit capture is affected by
varying price sensitivity $\alpha$, blended rate $P_0$, and non-buying
market share $s_0$, respectively.  Each data point in the figures is
obtained by varying each parameter over a range of values.  We vary
$\alpha$ between 1 and 10, $P_0$ between 5 and 30, and $s_0$ between 0
and 0.9. As we vary the parameters, we select and plot the {\em minimum}
observed profit capture over the whole parameter range, for the
profit-weighted strategy with different numbers of bundles. In other
words, these plots show the worst case relative profit capture for the
ISP over a range of parameter values.
The trend of these minimum profit
capture points is qualitatively similar to patterns in
Figures~\ref{fig:ced-division-types} and~\ref{fig:logit-division-types}.
For example, using the CED model and grouping flows in two bundles in the EU
ISP yields around 0.8 profit capture, regardless of price sensitivity,
blending rate, and market share. These results indicate that
our model is robust to a wide range of parameter values.

\section{Implementing Tiered Pricing}
\label{sec:implementation}

ISPs can implement
the type of tiered pricing that we describe in Section~\ref{sec:eval}
without any changes to their existing protocols or infrastructure, and
ISPs may already be using the techniques we describe below.  If that is
the case, they could simply apply a profit-weighted bundling strategy to
re-factor their pricing to improve their profit, possibly without even
making many changes to the network configuration. We describe two tasks
associated with tiered pricing: associating flows with tiers and
accounting for the amount of traffic the customer sends in each tier.

\subsection{Associating Flows with Tiers}

Associating each flow (or destination) with a tier can be done within the
context of today's routing protocols.  When the upstream ISP sends routes
to its customer, it can ``tag'' routes it announces with a label that
indicates which tier the route should be associated with; ISPs can use
BGP extended communities to perform this tagging. Because the communities
propagate with the route, the customer can establish routing policies on
every router within its own network based on these tags.

Suppose that a large transit service provider has routers in different
geographic regions.  Routers at an exchange point in, say, New York,
might advertise routes that it learned in Europe with a special tag
indicating that the path the route takes is trans-Atlantic and, hence, bears
a higher price than other, regional routes. The customer can then use the
tag to make routing decisions. For example, if a route is tagged as an
expensive long-distance route, the customer might choose to use its own
backbone to get closer to destination instead of performing the default
``hot-potato'' routing (\ie, offloading the traffic to a transit network
as quickly as possible). A large customer might also use this pricing
information to better plan its own network growth. 

\subsection{Accounting}
\label{sec:accounting}

Implementing tiered pricing requires accounting for traffic either on a
{\em per-link} or {\em per-flow} basis.

\begin{figure}[t]
  \centering
\subfigure[SNMP-Based
Accounting.\label{fig:multi-link}]{\includegraphics[width=0.6\linewidth]{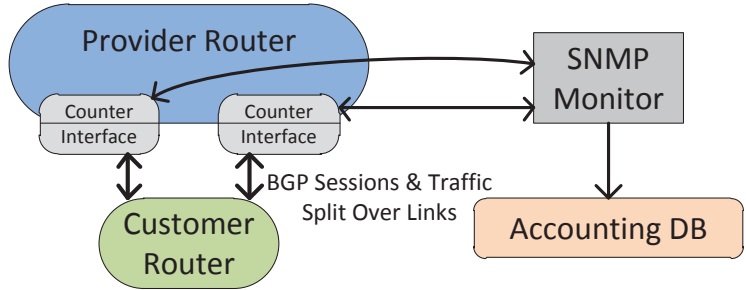}}
\subfigure[Flow-Based
Accounting.\label{fig:flow-based}]{\includegraphics[width=0.6\linewidth]{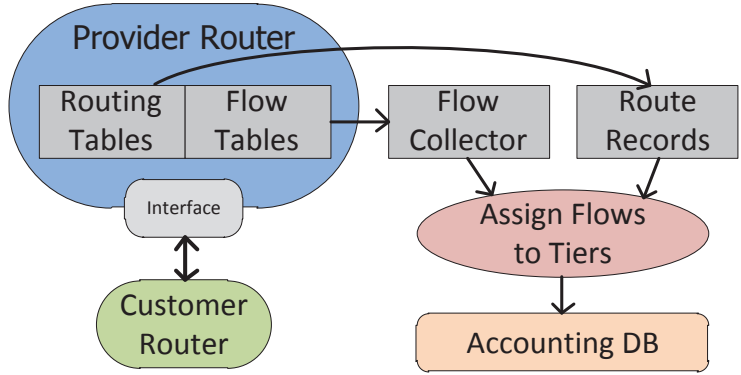}}
  \caption{Implementing accounting for tiered pricing.}
\end{figure}

\para{Link-Based Accounting.}  As shown in Figure~\ref{fig:multi-link},
an edge router can establish two or more physical or virtual links to
the customer, with a Border Gateway Protocol (BGP)~\cite{rfc4271}
session for each physical or virtual link. In this setup, each pricing
tier would have a separate link.  Each link carries the traffic only to the
set of destinations advertised over that session (\eg{}, on-net traffic,
backplane peering traffic). Because each link has a separate routing
session and only exchanges routes associated with that pricing tier, the
customer and provider can ensure that traffic for each tier flows over
the appropriate link: The customer knows exactly which traffic falls into
which pricing tier based on the session onto which it sends traffic.
Billing may also be simpler and easier to understand, since, in this
mode, a provider can simply bill each link at a different rate.
Unfortunately, the overhead of this accounting method grows significantly
with the number of pricing levels ISP intends to support.


%

\para{Flow-Based Accounting.} In flow-based accounting, as in traditional
peering and transit, an upstream ISP and a customer establish a link with
a single routing session. As shown in Figure~\ref{fig:flow-based}, the
accounting system collects both flow statistics (\eg, using
NetFlow~\cite{www-netflow}) and routing information to determine resource
usage. For the purposes of accounting, bundling effectively occurs after
the fact: flows can be mapped to distances using the routing table
information and priced accordingly, exactly as we did in our evaluation
in Section~\ref{sec:eval}. Assuming flow and routing information
collection infrastructure in place, flow-based accounting may be easier
to manage, and it is easier to bundle flows into different bins according
to various bundling strategies (\eg, profit-weighted bundling) {\em post
facto}.

\section{Related Work}\label{sec:related}

Developing and analyzing pricing models for the Internet is
well-researched in both networking and economics. Two aspects are 
most relevant for our work: the unbundling of connectivity and the
dimensions along which to unbundle it. Although similar studies of pricing
exist, none evaluated in the context of real network demand and
topology data.

The unbundling of connectivity services refers to the setting of
different prices for such services along various usage dimensions such as
volume, time, destination, or application type. Seminal works by Arrow
and Debreu~\cite{Arrow:1954} and McKenzie~\cite{McKenzie:1959} show that
markets where commodities are sold at infinitely small
granularities are more efficient.
More recent studies however, demonstrate that unbundling may be
inefficient in certain settings, such as when selling information goods
with zero or very low marginal cost (such as access to online
information)~\cite{Bakos1998, Laffont2003, Odlyzko:2011}.  This is not
always the case with the connectivity market, where ISPs incur different
costs to deliver traffic to different destinations.  In addition, many
service providers already use price discrimination~\cite{Odlyzko:2009}.

Kesidis {\em et al.}~\cite{Kesidis:2008} and Shakkottai {\em et
al.}~\cite{Shakkottai:2008} study the benefits of pricing connectivity
based on volume usage and argue that, with price differentiation, one can
use resources more efficiently. In particular, Kesidis {\em et al.} show
that usage-based unbundling may be even more beneficial to access
networks rather than core networks. 
Time is another dimension along which providers can unbundle
connectivity. Jiang {\em et al.}~\cite{Jiang:2008} study the role of time
preference in network prices and show analytically that service providers
can achieve maximum revenue and social welfare if they differentiate
prices across users and time. Hande {\em et al.}~\cite{Hande:2010}
characterize the economic loss due to ISP inability or unwillingness to
price broadband access based on time of day.

\section{Conclusion}\label{sec:conclusion}


As the price of Internet transit drops, transit providers are selling
connectivity using ``tiered'' contracts based on traffic cost,
volume, or destination to maintain profits. 
We have studied two questions: How does tiered pricing benefit the ISPs?
and How does tiered pricing affect their customers?
We developed a model for an Internet transit market that helps us to
evaluate differing pricing strategies of the ISPs.
We have applied our model to traffic demand and topology data from three
large ISPs to evaluate various bundling strategies.

We find that the ISPs gain most of the profits with {\em only three or
four pricing tiers} and likely have little incentive to icnrease
granularity of pricing even further. We also show that consumer surplus
follows closely, if not precisely, the increases in ISP profit with more
pricing tiers.

\begin{small}

\bibliographystyle{abbrv}

\balance

\bibliography{ref,rfc}

\begin{thebibliography}{10}

\bibitem{www-adam-unmetered}
{Adam Internet}.
\newblock Unmetered content.
\newblock \url{http://www.adam.com.au/unmetered/unmetered.php}.
\newblock Retrieved: June 2011.

\bibitem{Ali08}
G.~Ali.
\newblock Expected utility in econometric random utility models.
\newblock {\em The Indian Journal of Statistics}, 70(1), 2008.

\bibitem{Arrow:1954}
K.~J. Arrow and G.~Debreu.
\newblock Existence of equilibrium for competitive economy.
\newblock {\em Econometrica}, 22(3):265--290, 1954.

\bibitem{Bakos1998}
Y.~Bakos and E.~Bronjolfsson.
\newblock Bundling and competition on the {I}nternet.
\newblock {\em Marketing Science}, 19(1), Jan. 1998.

\bibitem{Besanko:logit1998}
D.~Besanko, S.~Gupta, and D.~Jain.
\newblock Logit demand estimation under competitive pricing behavior: An
  equilibrium framework.
\newblock {\em Management Science}, 44:1533--1547, Nov. 1998.

\bibitem{www-bsnl-pricelist}
BNSL.
\newblock Tariff for leased lines.
\newblock \url{http://www.bsnl.co.in/service/2mbps.pdf}.
\newblock Retrieved: June 2011.

\bibitem{telia-cogent-depeering}
M.~A. Brown, A.~Popescu, and E.~Zmijewski.
\newblock {Peering Wars: Lessons Learned from the Cogent-Telia De-peering}.
\newblock In {\em NANOG 43}, June 2008.

\bibitem{Chang:infocom2006}
H.~Chang, S.~Jamin, and W.~Willinger.
\newblock To peer or not to peer: Modeling the evolution of the {I}nternet's
  {AS}-level topology.
\newblock In {\em Proc. {IEEE INFOCOM}}, Barcelona, Spain, Mar. 2006.

\bibitem{www-chunghwa-pricelist}
{Chunghwa Telecom}.
\newblock Leased line pricelist.
\newblock \url{http://www.cht.com.tw/CHTFinalE/Web/Business.php?CatID=476}.
\newblock Retrieved: June 2011.

\bibitem{www-cisco-transceivers}
{Cisco}.
\newblock {SFP} optics for gigabit ethernet applications.
\newblock
  \url{http://www.cisco.com/en/US/prod/collateral/modules/ps5455/ps6577/product_data_sheet0900aecd8033f885.html}.
\newblock Retrieved: June 2011.

\bibitem{Dhamdhere2010}
A.~Dhamdhere, P.~Francois, and C.~Dovrolis.
\newblock A value based framework for internet peering agreements.
\newblock In {\em Proc. International Teletraffic Congress (ITC)}, 2010.

\bibitem{www-etisalat-pricelist}
{Etisalat ISP}.
\newblock Leased circuit rental charges.
\newblock \url{http://tinyurl.com/66tfvj6}.
\newblock Retrieved: June 2011.

\bibitem{Feamster2004c}
N.~Feamster and H.~Balakrishnan.
\newblock Verifying the correctness of wide-area {I}nternet routing.
\newblock Technical Report MIT-LCS-TR-948, Massachusetts Institute of
  Technology, May 2004.

\bibitem{Feamster2004b}
N.~Feamster, Z.~M. Mao, and J.~Rexford.
\newblock Border{G}uard: Detecting cold potatoes from peers.
\newblock In {\em Proc. {I}nternet Measurement Conference}, Taormina, Italy,
  Oct. 2004.

\bibitem{www-guavus-tiers}
{Guavus}.
\newblock Profitable tiered pricing.
\newblock \url{http://www.guavus.com/solutions/tiered-pricing}.
\newblock Retrieved: June 2011.

\bibitem{Hande:2010}
P.~Hande, M.~Chiang, R.~Calderbank, and J.~Zhang.
\newblock Pricing under constraints in access networks: Revenue maximization
  and congestion management.
\newblock In {\em Proc. {IEEE INFOCOM}}, San Diego, {CA}, Mar. 2010.

\bibitem{www-itu-handbook}
{ITU} {T}elecommunication {I}ndicator {H}andbook.
\newblock
  \url{http://www.itu.int/ITU-D/ict/publications/world/material/handbook.html}.
\newblock Retrieved: June 2011.

\bibitem{Jiang:2008}
L.~Jiang, S.~Parekh, and J.~Walrand.
\newblock Time-dependent network pricing and bandwidth trading.
\newblock In {\em Proc. IEEE BoD}, 2008.

\bibitem{Johari2003}
R.~Johari and J.~Tsitsiklis.
\newblock {Routing and Peering in a Competitive Internet}.
\newblock Technical Report LIDS Publication 2570, Massachusetts Institute of
  Technology, 2003.
\newblock \url{http://www.stanford.edu/~rjohari/pubs/netgamepaper.pdf}.

\bibitem{Kesidis:2008}
G.~Kesidis, A.~Das, and G.~D. Veciana.
\newblock On flat-rate and usage-based pricing for tiered commodity {I}nternet
  services.
\newblock In {\em Proc. CISS}, 2008.

\bibitem{Labovitz:sigcomm2010}
C.~Labovitz, S.~Iekel-Johnson, D.~McPherson, J.~Oberheide, and F.~Jahanian.
\newblock Internet inter-domain traffic.
\newblock In {\em {Proc.\ ACM SIGCOMM}}, New Delhi, India, Aug. 2010.

\bibitem{Laffont2003}
J.-J. Laffont, S.~Marcus, P.~Rey, and J.~Tirole.
\newblock Internet interconnection and the off-net-cost principle.
\newblock {\em The Rand Journal of Economics}, 34(2), 2003.

\bibitem{www-he-cogent-depeering}
M.~Leber.
\newblock {IPv6} {I}nternet broken, cogent/telia/hurricane not peering.
\newblock http://www.merit.edu/mail.archives/nanog/msg01006.html, Oct. 2009.

\bibitem{www-comcast-vs-level3}
{Level 3 Communications}.
\newblock {L}evel 3 statement concerning {C}omcast's actions.
\newblock \url{http://www.level3.com/index.cfm?pageID=491&PR=962}.
\newblock Retrieved: June 2011.

\bibitem{www-geoip}
{MaxMind GeoIP Country}.
\newblock \url{http://www.maxmind.com/app/geolitecountry}.
\newblock Retrieved: June 2011.

\bibitem{McFadden1973}
D.~McFadden.
\newblock Conditional logit analysis of qualitative choice behavior.
\newblock {\em Frontiers Of Econometrics}, 1974.

\bibitem{McKenzie:1959}
L.~W. McKenzie.
\newblock On the existence of general equilibrium for a competitive economy.
\newblock {\em Econometrica}, 1959.

\bibitem{Mo:2000}
J.~Mo and J.~Walrand.
\newblock Fair end-to-end window-based congestion control.
\newblock {\em IEEE/ACM Trans. Netw.}, 8(5):556--567, 2000.

\bibitem{Odlyzko:2011}
P.~Nabipay, A.~M. Odlyzko, and Z.-L. Zhang.
\newblock Flat versus metered rates, bundling, and ``bandwidth hogs''.
\newblock In {\em 6th Workshop on the Economics of Networks, Systems, and
  Computation}, San Jose, CA, June 2011.

\bibitem{www-netflow}
Cisco {N}et{F}low.
\newblock
  {\url{http://www.cisco.com/en/US/products/ps6601/products_ios_protocol_group_home.html}}.
\newblock Retrieved: June 2011.

\bibitem{Norton-www}
W.~Norton.
\newblock {D}r{P}eering.net.
\newblock \url{http://drpeering.net}.

\bibitem{www-ntt-pricelist}
{NTT East}.
\newblock Leased circuit price list.
\newblock \url{http://www.ntt-east.co.jp/senyo_e/charge/digital.html}.
\newblock URL retrieved January 2011.

\bibitem{Odlyzko:2009}
A.~M. Odlyzko.
\newblock Network neutrality, search neutrality, and the never-ending conflict
  between efficiency and fairness in markets.
\newblock {\em {Review of Network Economics}}, 8(1):40--60, Mar. 2009.

\bibitem{www-ore-pricelist}
{ORE}.
\newblock Wholesale leased lines price list.
\newblock
  \url{http://www.otewholesale.gr/Portals/0/LEASED%20LINES_Pricelist_ENG_081110.pdf}.
\newblock Retrieved: June 2011.

\bibitem{peeringdb}
{Peering Database}.
\newblock \url{http://www.peeringdb.com}.

\bibitem{l3-cogent-depeering}
A.~Popescu and T.~Underwood.
\newblock {D(3) Peered: Just the Facts Ma¿am. A Technical Review of Level
  (3)¿s Depeering of Cogent}.
\newblock In {\em NANOG 35}, Oct. 2005.

\bibitem{rfc4271}
Y.~Rekhter, T.~Li, and S.~Hares.
\newblock {\em {A Border Gateway Protocol 4 (BGP-4)}}.
\newblock {Internet Engineering Task Force}, Jan. 2006.
\newblock RFC 4271.

\bibitem{Shakkottai:2008}
S.~Shakkottai, R.~Srikant, A.~E. Ozdaglar, and D.~Acemoglu.
\newblock The price of simplicity.
\newblock {\em IEEE Journal on Selected Areas in Communications},
  26(7):1269--1276, 2008.

\bibitem{www-telegeography-prices}
{Telegeography}.
\newblock Bandwidth pricing report.
\newblock
  \url{http://www.telegeography.com/product-info/pricingdb/download/bpr-2009-10.pdf}.
\newblock Retrieved: June 2011.

\end{thebibliography}

\label{lastpage}

\end{small}

\end{sloppypar}
\end{document}